\documentclass[12pt,a4paper]{article}
\usepackage{graphicx}
\usepackage{amssymb,amsfonts,amsmath}
\usepackage{cite}
\renewcommand{\citeleft}{(}
\renewcommand{\citeright}{)}
\renewcommand{\citepunct}{; }

\makeatletter
\renewcommand{\@biblabel}[1]{}
\makeatother

\newcommand{\EQ}{Eq.~}
\newcommand{\EQS}{Eqs.~}
\newcommand{\FIG}{Fig.~}
\newcommand{\FIGS}{Figs.~}
\newcommand{\TAB}{Tab.~}

\newcommand{\SEC}{Sec.~}
\newcommand{\SECS}{Secs.~}

\begin{document}

\title{Directionality of contact networks suppresses selection pressure
in evolutionary dynamics}
\bigskip
\bigskip
\author{Naoki Masuda\\
\ \\
\ \\
Graduate School of Information Science and Technology,\\
The University of Tokyo,\\
7-3-1 Hongo, Bunkyo, Tokyo 113-8656, Japan\\
\ \\
PRESTO, Japan Science and Technology Agency,\\
4-1-8 Honcho, Kawaguchi, Saitama 332-0012, Japan}

\setlength{\baselineskip}{0.77cm}
\maketitle
\newpage

\begin{abstract}
\setlength{\baselineskip}{0.77cm}
Individuals of different types, may it be genetic, cultural, or else, with different levels of fitness often compete for reproduction and survival. A fitter type generally has higher chances of disseminating their copies to other individuals. The fixation probability of a single mutant type introduced in a population of wild-type individuals quantifies how likely the mutant type spreads.  How much the excess fitness of the mutant type increases its fixation probability, namely, the selection pressure, is important in assessing the impact of the introduced mutant. Previous studies mostly based on undirected and unweighted contact networks of individuals showed that the selection pressure depends on the structure of networks and the rule of reproduction. Real networks underlying ecological and social interactions are usually directed or weighted. Here we examine how the selection pressure is modulated by directionality of interactions under several update rules. Our conclusions are twofold. First, directionality discounts the selection pressure for different networks and update rules. Second, given a network, the update rules in which death events precede reproduction events significantly decrease the selection pressure than the other rules.
\end{abstract}


\newpage

\section{Introduction}\label{sec:introduction}

In evolutionary dynamics, different types of individuals compete for
survival in a population.  A type means genotype, social behavior,
cultural trait, and so on,
depending on the context. A type that is fitter than others generally
bears more offsprings. A major
quantity representing how successfully a type spreads in evolutionary
dynamics is the fixation probability
\cite{Moran58,Ewensbook,Lieberman05,Nowak06book}.  In a
simple case in which
there are only two types, the fixation probability of a type is the
probability that a single individual of that mutant type introduced in a
population of the other wild-type individuals eventually occupies the
entire population. Requirements for considering the fixation
probability are that the evolutionary dynamics are stochastic and that
the two unanimity states, that is, the one of the introduced type and
the other of the wild type, are the only two absorbing states.  The
fixation probability of a type depends on the
fitness of the type, connectivity of individuals, and the update rule of
evolutionary dynamics 
\cite{Ewensbook,Lieberman05,Antal06prl,Nowak06book,Sood08pre}.

Evolutionary dynamics, both ecological and social, pretty often occur
on complex contact networks of individuals
\cite{Newman03siam,Watts04,Keeling05inter,Proulx05tee,May06tee}.  Some
networks as well as the rule of reproduction and other factors
amplify the selection
pressure in the sense that a fitter type has a larger fixation
probability and a less fit type has a smaller fixation probability
compared to a reference case of the all-to-all connected population
\cite{Lieberman05,Antal06prl,Sood08pre}.  Other combinations of a
network and an update rule may suppress evolutionary pressure, with
the fixation probability relatively insensitive to the
fitness of the mutant type.  To quantify the extent to which a
particular situation amplifies or suppresses the selection pressure is
important for assessing the impact of a mutant type.

Studies of fixation probability on networks and structured populations
have been restricted to two neutral types
\cite{Donnelly83,Taylor90,Taylor96} or undirected and unweighted
networks \cite{Maruyama70,Slatkin81,Antal06prl,Sood08pre}, albeit a
notable exception \cite{Lieberman05}. Regarding the directionality,
undirected (and unweighted) networks may be natural for modeling
social interaction such as games, where an adjacent pair of
individuals, for example, is simultaneously involved in a single game
to determine the fitness of each individual.  However,
reproduction events given the fitness of each individual
may occur on directed or
weighted networks.  Indeed, relevant contacts in many real networks
have directionality because of, for example, heterogeneity in the size
of habitat patches \cite{Gustafson96} and geographical biases such as
the wind direction \cite{Schooley03} and riverine streams
\cite{Schick07} in ecological networks.  Social networks based on
glooming behavior of rhesus monkeys \cite{Sade72}, email communication
of humans \cite{Ebel02pre_email,Newman02pre_email} also have
directionality. In the present work, we examine the evolutionary
dynamics on different undirected and directed networks, namely, the
complete graph, the undirected cycle, the directed cycle, the weighted
undirected star, undirected and directed random graphs, and undirected
and directed scale-free (SF) networks.  Motivated by previous studies
\cite{Antal06prl,Ohtsuki06nat,Ohtsuki06jtb,Sood08pre}, we examine the
effects of several update rules on the selection pressure.  We show
that asymmetric connectivity generally turns down the selection
pressure and that specific update rules suppress the selection
pressure more than other rules.

\section{Model}\label{sec:model}

\subsection{Dynamics}

We consider a population of $N$ haploid, asexually reproducing
individuals.  The structure of the population is described by a
directed graph $G$.  We denote by $E$ the set of directed edges,
implying that
$(v_i,v_j)\in E$ if and only if there is a
directed edge from $v_i$ to $v_j$. Each
node $v_i$ ($1\le i\le N$) is occupied by an individual of
one type. Edges
represent the likelihood with which the types
are transferred from nodes to nodes in evolutionary dynamics.

We assume that there are two types $A$ and $B$.
A node is occupied by either type $A$ or $B$.
Types $A$ (mutant type) and $B$ (wild type)
confer fitness $r$ and 1 to its bearer, respectively.
Given graph $G$ and an update rule, 
the collective network state, which is specified by
an assignment of $A$ or $B$ to each node of the network,
stochastically evolves.
The dynamics last until the unanimity (i.e., fixation)
of $A$ or that of $B$ is
reached. We do not assume mutation so that 
these two collective states are the only absorbing states.
We are concerned with the fixation probability of type $A$ denoted by
$F_v(r)$. It is the probability that a single type-$A$ mutant introduced
at node $v$ in a sea of $N-1$ resident type-$B$ 
individuals fixates. Type $B$ fixates with probability $1-F_v(r)$.
The fixation probability depends on $G$,
the update rule, and the initial location $v$ of the $A$ mutant.
To focus on the effect of $G$ and the update rule,
we examine $F(r)=\sum_{v}F_v(r)/N$.
The effect of the 
initial location of the mutant has been analyzed
for undirected \cite{Antal06prl,Sood08pre}
and directed \cite{MasudaOhtsuki09} networks.
 
An update event occurs on one directed edge per unit time.
We assume that the direction of reproduction is the same as the edge
direction. We
examine the following five update rules, most of which are motivated
by past literature. For the sake of explanation, we
explain the update rules for directed and unweighted
networks.
However, we can extend the model and the results to
the case of weighted edges in a straightforward manner
\cite{Lieberman05} (also see related
analysis in \SEC\ref{sub:star}).
 
\subsection{BD-B update rule}\label{sub:BD-B}

Under the birth--death rule with selection on the birth (BD-B), we
first select one node $v_i$ for reproduction
in each time step. The probability that $v_i$ is selected
is proportional to the fitness value, that is, $f_i/\sum_l f_l$, where
$f_i\in \{r,1\}$ is the fitness of the type on $v_i$.
Note that selection operates on the
birth. Then, the type at $v_i$
is propagated to a neighbor of $v_i$ along a directed edge that
is chosen with probability $1/k_i^{out}$, where $k_i^{out}$ is
the outdegree of $v_i$.  The probability that directed edge
$(v_i,v_j)\in E$ is used for
reproduction is equal to $f_i/k_i^{out}\sum_l f_l$.

BD-B is the update rule considered in \cite{Lieberman05}.
It is equivalent to the previously defined birth--death rule for
games on undirected networks
\cite{Ohtsuki06nat,Ohtsuki06jtb} and to the invasion
process defined in \cite{Antal06prl,Sood08pre}. 

\subsection{BD-D update rule}

Under the birth--death rule with selection on the death (BD-D), a random
individual $v_i$ is first chosen for reproduction
with equal probability $1/N$. Then, one of its
neighbors $v_j$ that receives a directed
edge from $v_i$ dies with probability
proportional to $1/f_j$, and the type at $v_i$ replaces that at $v_j$.
Selection operates on the death.
The probability that edge $(v_i,v_j)\in E$ 
is used for reproduction is equal to
$f_j^{-1}/N\sum_{l; (v_i,v_l)\in E} f_l^{-1}$.
When $r=1$, BD-B and BD-D are identical.

\subsection{DB-B update rule}

Under the death--birth rule with selection on the birth (DB-B), a
random individual $v_j$ first dies with equal probability $1/N$. Then,
a neighbor of $v_j$ that sends a directed edge to $v_j$,
denoted by $v_i$, is selected for reproduction
with probability $f_i/\sum_{l; (v_l,v_j)\in E} f_l$, and
the type of $v_i$ replaces that at $v_j$. Selection
operates on the birth. The
probability that $(v_i,v_j)\in E$ is used for reproduction is equal to
$f_i/N\sum_{l; (v_l,v_j)\in E}f_l$.  DB-B is equivalent to the
score-dependent fertility model proposed in \cite{Nakamaru98jtb} and
to the death--birth rule previously used for evolutionary games on
undirected networks \cite{Ohtsuki06nat,Ohtsuki06jtb}.
Many numerical studies of spatial reciprocity \cite{Nowak92} and
network reciprocity \cite{Santos05prl} are also based on this update
rule or similar rules.

\subsection{DB-D update rule}

Under the death--birth rule with selection on the death (DB-D),
a node $v_j$ is first chosen for death
with probability
$f_j^{-1}/\sum_l f_l^{-1}$. 
Selection operates on the death.
Next, a neighbor of $v_j$ that sends a directed edge
to $v_j$, denoted by $v_i$,
is chosen for reproduction randomly with probability $1/k_j^{in}$, where
$k_j^{in}$ is the indegree of $v_j$.
Then, the type at $v_i$ replaces that at $v_j$. 
The probability that $(v_i,v_j)\in
E$ is used for reproduction 
is equal to $f_j^{-1}/k_j^{in}\sum_l f_l^{-1}$.
This rule is equivalent to the score-dependent viability model
proposed in \cite{Nakamaru97jtb} and to
the voter model defined in \cite{Antal06prl,Sood08pre}.
When $r=1$, DB-B and DB-D are identical.

\subsection{LD update rule}\label{sub:LD}

Under the link dynamics (LD) with selection on the birth (resp. death),
directed edge $(v_i,v_j)\in E$ is chosen 
with probability $f_i/\sum_{(v_l,v_{l^{\prime}})\in E}f_l$
(resp. $f_j^{-1}/\sum_{(v_{l^{\prime}},v_l)\in E}f_l^{-1}$).
The denominators in these probabilities indicate the
summation over all the directed edges.
Selection operates on the birth (resp. death). 
Then, the type at $v_i$ replaces that at $v_j$.
When there are only two types, as assumed in this and most previous 
studies \cite{Lieberman05,Antal06prl,Sood08pre},
the LD with selection on the birth and that on the death
coincide with each other, up to a change of
the timescale. Then
this rule is the same as
the LD defined in \cite{Antal06prl,Sood08pre}.

\section{Evolutionary amplifiers and suppressors}

In this section, we explain the order parameter measured in the
following 
analysis.  We are concerned with dependence of selection pressure on
networks and update rules.  The fixation probability at
neutrality, namely, $r=1$, is equal to $F(1)=1/N$.  A type with a
larger fitness is more likely to fixate, that is, $dF(r)/dr\ge 0$.
However, how $F(r)$ depends on $r$ differs by the network and by the
update rule.  The reference profile of $F(r)$ is given by the Moran
process \cite{Moran58,Ewensbook,Nowak06book}. In the Moran
process, we select an individual $v_i$ at each time step for
reproduction with the probability proportional to $f_i$. Then, the
offspring replaces an individual randomly picked from the rest of the
population with the equal probability $1/(N-1)$. The evolutionary
dynamics on the complete graph under BD-B is the Moran
process.  As explained in Appendix~B, $F(r)$ for the Moran process is
equal to $F(r)$ for the complete graph under DB-D and LD as well as
under BD-B, and it is
given by
\begin{equation}
\frac{1-r^{-1}}{1-r^{-N}}.
\label{eq:F moran}
\end{equation}
A network or an update rule
that yields $F(r)$ larger than \EQ\eqref{eq:F moran} for $r>1$ and
smaller than \EQ\eqref{eq:F moran} for $r<1$ is called evolutionary
amplifier. For an evolutionary amplifier,
the selection pressure is magnified relative to that of the Moran process
\cite{Lieberman05}.
If the fixation probability is
smaller than \EQ\eqref{eq:F moran} for $r>1$ and larger than
\EQ\eqref{eq:F moran} for $r<1$, the network or the update rule
is called evolutionary suppressor.

The amplification factor $K$
may be defined when the fixation probability is given in the form:
\begin{equation}
\frac{1-r^{-1}}{1-r^{-KN}}.
\label{eq:F K}
\end{equation}
If \EQ\eqref{eq:F K} holds for a graph with $N$ nodes,
the evolutionary dynamics are equivalent to 
the Moran process with the effective population size
$KN$. Because a larger population size results in
stronger selection pressure
for the Moran process,
$K>1$ and $K<1$ correspond to evolutionary
amplifier and suppressor, respectively.
For example, for a family of star graphs under BD-B
\cite{Lieberman05}, and for undirected uncorrelated graphs
under BD-B and DB-D \cite{Antal06prl,Sood08pre},
$F(r)$ obeys \EQ\eqref{eq:F K} with different values of $K$
in an appropriate limit.
However, $F(r)$ generally
deviates from \EQ\eqref{eq:F K}. Therefore, in the numerical
simulations in \SEC\ref{sec:numerical results}, we calculate
the fixation probability at $r=4$ and compare it with that of the Moran
process.

\section{Fixation probability for some simple graphs}\label{sec:simple}

For some simple graphs, we calculate the fixation probability for
different update rules to determine the selection pressure.
The details of the calculations
are shown in Appendix~B.
The results in this section
are mostly restricted to undirected networks. We will
treat the effect of directionality numerically
in \SEC\ref{sec:numerical results}.

\subsection{Complete graph}\label{sub:complete}

Consider the evolutionary dynamics on the complete graph of $N$ nodes
depicted in \FIG\ref{fig:graphs}(a). Self loops are excluded.  The
complete graph is undirected and regular, where a regular graph is one
in which the indegree and the outdegree of all the nodes are the
same. For undirected regular graphs, BD-B, DB-D, and LD are all
equivalent to the Moran process with the same population size $N$
\cite{Antal06prl,Sood08pre}. Therefore, $F(r)$ is given by
\EQ\eqref{eq:F moran}.
For BD-D, we obtain
\begin{equation}
F(r)=\frac{1}{1+(N-1)\sum^{N-1}_{m=1}\frac{r^{-m+1}}{m+r(N-1-m)}}.
\label{eq:F BD-D complete}
\end{equation}
For DB-B, we obtain
\begin{equation} F(r)
=\frac{N-1}{N}\frac{1-r^{-1}}{1-r^{-N+1}}.
\label{eq:F DB-B complete}
\end{equation}

In \FIG\ref{fig:F simple}(a),
$F(r)$ for the Moran process, BD-D, and DB-B is compared
for the complete graph with $N=10$.
The three lines cross at neutrality, that
is, $(r,F(r))=(1,1/N)$.
The complete graph is an evolutionary suppressor under BD-D,
even though $\lim_{r\to\infty}F(r)=1$ and
$\lim_{r\to 0}F(r)=0$.
This is because selection occurs among the $N-1$ nodes
excluding the reproducing node.
DB-B is a stronger suppressor than BD-D.
Differently from BD-D,
we obtain $\lim_{r\to\infty}F(r)=(N-1)/N$
and $\lim_{r\to 0}F(r)=0$ for DB-B, so that
a mutant may not fix under DB-B even if its fitness is
infinitely large.

For the three different cases, $F(4)$ is plotted against $N (\ge 2)$
in \FIG\ref{fig:F simple}(b).  The fixation probability for BD-D and
DB-B converges to that for the Moran process as $N\to\infty$.  
Remarkably,
for BD-D and DB-B, the existence of more competitors in the
population, that is, larger $N$, leads to the higher fixation
probability of the single type-$A$ mutant.
As $N$ increases, selection acts on
more nodes relative to the population size (i.e., $(N-1)/N$)
under BD-D and DB-B.  Then type-$A$ nodes, which are
rare in an early stage of dynamics, are involved in
competition for survival (under BD-D) or reproduction (under DB-B)
more often such that type $A$ takes advantage of being inherently
fitter than type $B$.

\subsection{Undirected cycle}\label{sub:cycles}

Consider the undirected cycle of size $N$ depicted in
\FIG\ref{fig:graphs}(b).
Because the undirected cycle is unweighted and
regular, BD-B, DB-D, and LD are again equivalent to
the Moran process, and
$F(r)$ is given by \EQ\eqref{eq:F moran}.
For BD-D, we obtain
\begin{equation}
F(r)= \frac{1-r^{-1}}{1+\frac{r-1}{r(r+1)}
+\frac{(r^2-2r-1)r^{-N+1}}{r+1}}.
\label{eq:F BD-D cycle}
\end{equation}
For DB-B, we obtain
\begin{equation}
F(r)= \frac{1-r^{-1}}{1+\frac{r-1}{2r}
+\frac{(r-3)r^{-N+1}}{2}}.
\label{eq:F DB-B cycle}
\end{equation}

For the undirected cycle, $F(r)$
with $N=10$ and
$F(4)$ with $N$ varied are shown in
\FIG\ref{fig:F simple}(c) and
\FIG\ref{fig:F simple}(d), respectively.
Qualitatively agreeing with the case of the complete graph,
BD-D is suppressing, and DB-B is even more so.
In contrast to the case of the complete
graph, BD-D and DB-B persist to be suppressing for large $N$.
Under BD-D and DB-B, selection operates on about $\left<k\right>$
individuals, where $\left<k\right>$ is the mean degree of the network, 
whereas selection operates on $N$ individuals under
BD-B, DB-D, and LD.
For the complete graph,
the difference diminishes as $N\to\infty$ because
$\left<k\right>=N-1$.
For the undirected cycle, $\left<k\right>=2$ independent of $N$,
which is a likely reason why BD-D and DB-B are suppressing
even for large $N$.

\subsection{Directed cycle}

Consider the directed cycle of size $N$ depicted
in \FIG\ref{fig:graphs}(c).
It is straightforward to verify that the evolutionary dynamics under
BD-B, DB-D, and LD
are equivalent to the Moran
process. For BD-D and DB-B, selection pressure is totally annihilated, 
that is, $F(r)=1/N$.

\subsection{Star}\label{sub:star}

Consider the star with $N$ nodes depicted in \FIG\ref{fig:graphs}(d).
One central hub is connected to the other $N-1$ leaves. Only in this
section, we introduce the edge weight for a computation
purpose. Specifically, each edge outgoing from the hub has weight $1$,
and each edge incoming to the hub has weight $a$.  The edge weight is
assumed to be multiplied to the likelihood with which the edge is used
for reproduction (see \SECS\ref{sub:BD-B}-\ref{sub:LD} and
Appendix~B).

The fixation probability for the weighted star under LD is given by
\begin{equation}
F(r)=
\frac{1-\frac{N-1}{N}\frac{r+a}{r(ra+1)}-\frac{1}{N}\frac{ra+1}{r(r+a)}}
{1-r^{-N}\left(\frac{r+a}{ra+1}\right)^{N-2}}.
\label{eq:star LD final}
\end{equation}
For general $a$, we obtain in the limit $N\to\infty$
\begin{equation}
K=-\frac{\ln \frac{r+a}{r(ra+1)}}{\ln r},
\label{eq:K star LD}
\end{equation}
where $K$ is defined by \EQ\eqref{eq:F K}.
Although in an incomplete form of \EQ\eqref{eq:K star LD}, which includes $r$ in the RHS,
$K>1$ ($K<1$) for all $r\neq 1$ when $a>1$ ($a<1$). Therefore,
the weighted star is an amplifier (a suppressor) when $a>1$ ($a<1$).
With $a=1$ in \EQ\eqref{eq:star LD final}, $F(r)$ is equal to
\EQ\eqref{eq:F moran}. This is expected because the
evolutionary dynamics for any undirected network
are equivalent to the Moran process under LD
\cite{Antal06prl,Sood08pre}. 

As shown in Appendix~B,
LD on the weighted star
with $a=N-1$ is equivalent to 
BD-B on the unweighted star.
Therefore, the unweighted star is an amplifier
under BD-B. Particularly,
in the limit $N\to\infty$, \EQ\eqref{eq:star LD final}
yields $K=2$, consistent with the previous result \cite{Lieberman05}.
The theory for undirected uncorrelated networks
\cite{Antal06prl,Sood08pre} also predicts
$K=\left<k\right>\left<k^{-1}\right>\approx 2$ for the undirected
star under BD-B.

Also shown in Appendix~B is that
LD on the weighted star with $a=1/(N-1)$ is equivalent to
DB-D on the unweighted star.
Setting $a=1/(N-1)$ in \EQ\eqref{eq:K star LD} yields
$K=0$ in the limit $N\to\infty$.
This is consistent with the result for
undirected uncorrelated networks: 
$K=\left<k\right>^2/\left<k^2\right>=2/N$
\cite{Antal06prl,Sood08pre}.

The fixation probability for BD-D is given by
\begin{equation}
F(r)=\frac{(N-1)N\frac{1+r(N-2)}{2+r(N-2)}+1}{N}
\frac{1}{1+\frac{1+r(N-1)}{r-1}
\left(1-\frac{r^{-N+2}N}{N-1+r}\right)},
\label{eq:star BD-D final}
\end{equation}
and that for DB-B is given by
\begin{equation}
F(r)=\frac{(rN+N+2r-2)(rN-r+1)}{N^2(N+2r-2)(r+1)}.
\label{eq:star DB-B final}
\end{equation}

\begin{figure}
\begin{center}
\includegraphics[height=4cm,width=6cm]{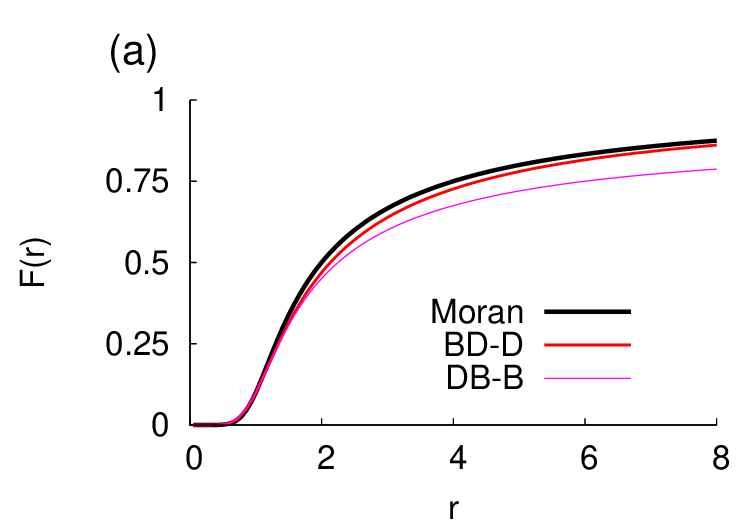}
\includegraphics[height=4cm,width=6cm]{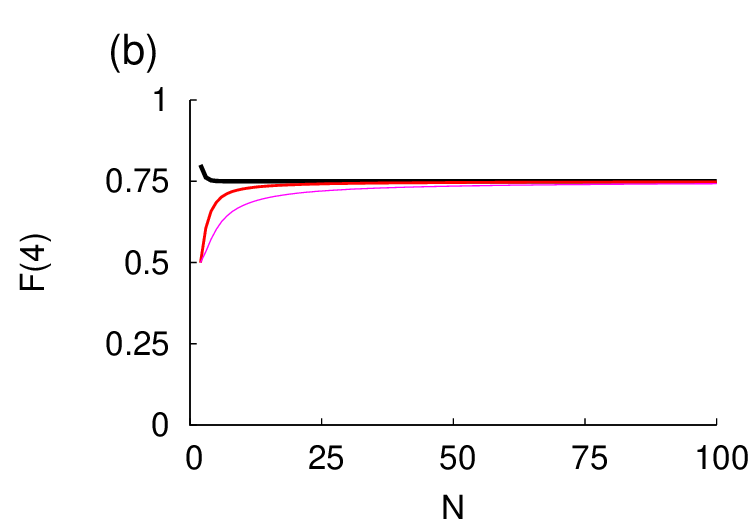}
\includegraphics[height=4cm,width=6cm]{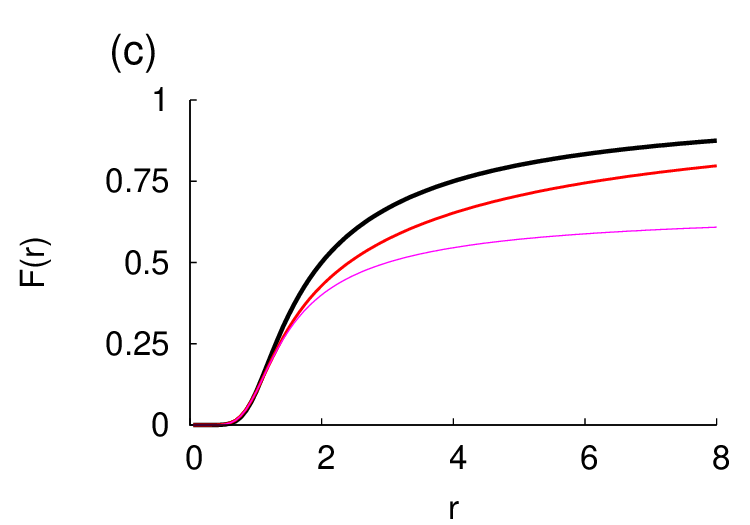}
\includegraphics[height=4cm,width=6cm]{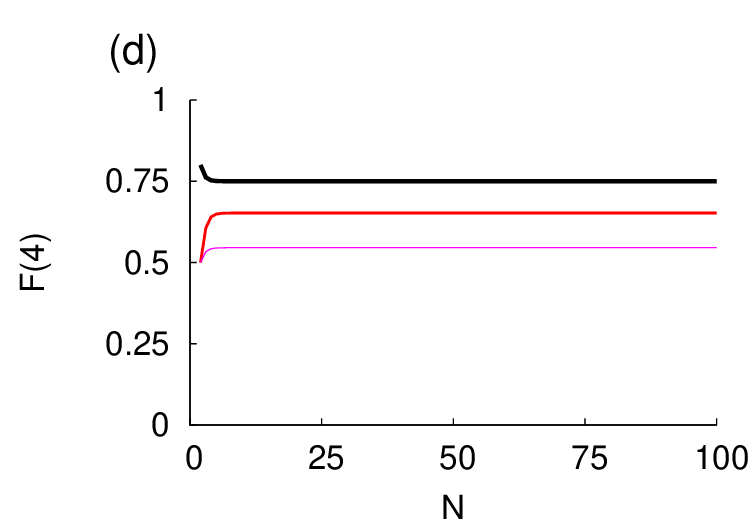}
\includegraphics[height=4cm,width=6cm]{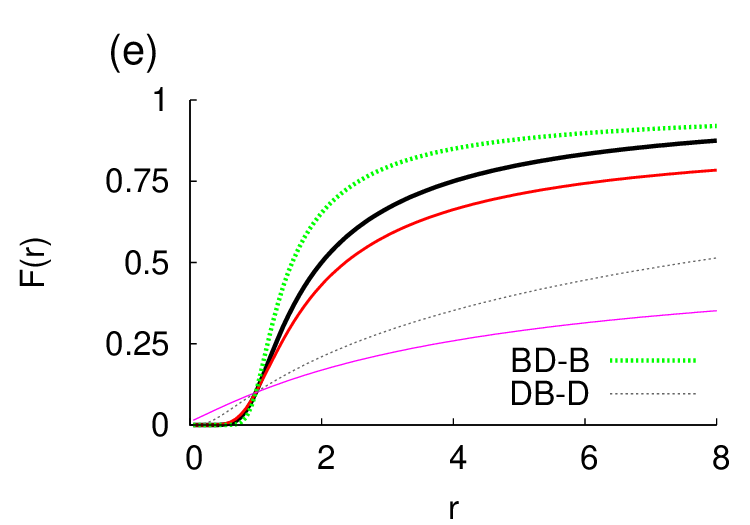}
\includegraphics[height=4cm,width=6cm]{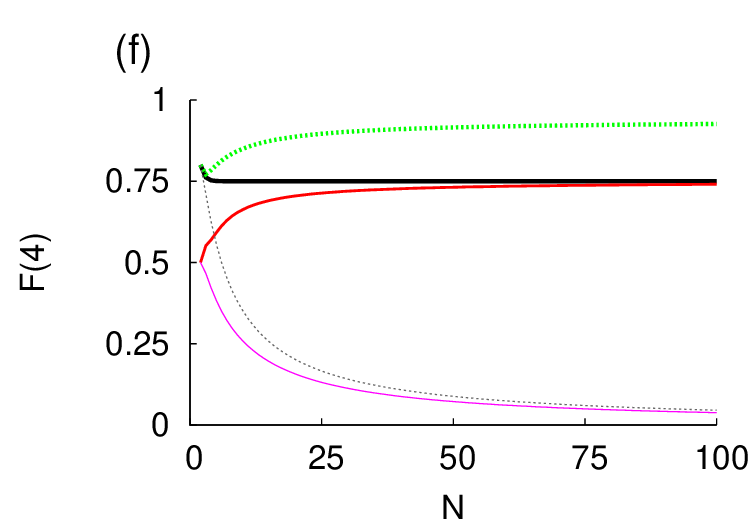}
\caption{(Color online)
Fixation probability for (a, b) the complete graph, (c, d)
the undirected cycle, and (e, f) the star, under different update rules.
We set $N=10$ and vary $r$ in (a, c, e). We set $r=4$ and vary $N$
in (b, d, f).
Thick solid black lines, medium solid red lines, and thin solid magenta
lines correspond to the standard Moran process,
BD-D, and DB-B, respectively. In (e, f),
thick dashed green lines and thin dashed gray lines correspond to
BD-B and DB-D, respectively.}
\label{fig:F simple}
\end{center}
\end{figure}

For the unweighted star, $F(r)$ with $N=10$ and $F(4)$ with $N$ varied
are shown in \FIG\ref{fig:F simple}(e) and \FIG\ref{fig:F simple}(f),
respectively. LD 
corresponds to the Moran process, and BD-B
is the only amplifying update rule. Among the other three
suppressing update rules, DB-B is the most suppressing.
Figure~\ref{fig:F simple}(f) indicates that BD-D is more
suppressing than DB-B for small $N$ and vice versa for large $N$.  As
$N$ tends large, DB-B and DB-D become strongly suppressing. Indeed,
both \EQ\eqref{eq:star LD final} with $a=1/(N-1)$ representing
DB-D and \EQ\eqref{eq:star DB-B final}
representing DB-B yield $F(r)=O(1/N)$, $r>1$.
In contrast, the fixation probability for BD-D approaches
that for the Moran process as $N\to\infty$; \EQ\eqref{eq:star BD-D
final} yields $\lim_{N\to\infty}F(r)=(r-1)/r$, which agrees with
the result
for the Moran process derived from \EQ\eqref{eq:F moran}.

\section{Numerical results}\label{sec:numerical results}

Here we 
report numerical results for the fixation probability of the mutant
type with $r=4$. 
Equation~\eqref{eq:F moran} implies
$\lim_{N\to\infty}F(r)=3/4$ for the Moran process.
If $F(4)$ measured for a combination of a network
and an update rule is larger (smaller) than $3/4$, that
combination is
probably amplifying (suppressing).
Whether the combination is amplifying or suppressing and to what extent
actually 
depend on $r$. However, our extensive numerical results suggest that
this dependence is not strong unless $r$ is extremely small or large. The
values of $F(4)$ are also well correlated with $dF(r)/dr|_{r=1}$,
which is the sensitivity of the selection pressure at neutrality.
Therefore, we assume $r=4$ in the following.

A graph $G$ is called strongly connected if there is a 
directed path from an arbitrary chosen node to another.
For $G$ that is not strongly connected, the mutant type introduced at
a node in a downstream component never fixates, and
the fixation problem is ascribed to that for the most upstream
strongly connected component of $G$
\cite{Lieberman05}. Therefore, we assume that $G$ is strongly connected
without loss of generality.

\subsection{Small networks}\label{sub:N=6}

In this section, we present numerical results for small networks. The
evolutionary dynamics are interpreted as a discrete-time Markov chain
on a finite space. A
state of the Markov chain is specified by an assignment of type $A$ or $B$
to each node, so that there are $2^N$ possible states.  Among them,
the two states corresponding to the unanimity of $A$ and that of $B$
are the unique absorbing states. Given $G$ and an update rule,
we obtain the exact fixation probability by solving a system of
linear equations of $2^N-2$ variables (see
Appendix~A for methods).  Because of the computation
time, we set $N=6$ and calculate $F(4)$ for all the 1047008 strongly
connected networks under the five update rules.  The results are
qualitatively similar for $N=4$ and $N=5$ (data not shown).

To visualize the results and to correlate $F(4)$ with the structure of $G$,
we regard the values of $F(4)$ for different
networks as different data points. Then we regress $F(4)$ of these
data points against various measurements of the network, which we call
order parameters. We look for the order parameters that are more
correlated with $F(4)$ than other order parameters.  Our choice of the order
parameters is arbitrary.  For each of the five update rules, we list
in \TAB\ref{tab:regressor} the Pearson correlation coefficient between
$F(4)$ and the order parameters, where $\left<\cdot\right>$ denotes
the average over the nodes. The first order parameter is the mean
degree $\left<k\right>\equiv
\left<k^{in}\right>=\left<k^{out}\right>$, where $k^{in}$ and
$k^{out}$ is the indegree and the outdegree of a node, respectively.
Seven order parameters are normalized moments of the degree
distribution: $\left<k\right>\left<1/k^{in}\right>$,
$\left<k\right>\left<1/k^{out}\right>$,
$\left<k\right>^2/\left<\left(k^{in}\right)^2\right>$,
$\left<k\right>^2/\left<\left(k^{out}\right)^2\right>$,
$\left<k\right>^2/\left<k^{in}k^{out}\right>$,
$\left<k^{out}/k^{in}\right>$, and $\left<k^{in}/k^{out}\right>$. The
other two order parameters are the normalized standard deviation of
the node temperature: $std(T^{in})$ and $std(T^{out})$.  The
temperature $T^{in}_i$ of node $v_i$ \cite{Lieberman05} is defined by
$T^{in}_i \equiv \sum_{j; (v_j,v_i)\in E}
\left(w_{ji}/\sum_l w_{jl}\right)$.
We define $std(T^{in})\equiv$
$\sqrt{\sum_{i=1}^N\left(T^{in}_i-\left<T^{in}\right>\right)^2/
N}/\left<T^{in}\right>$ $=
\sqrt{\sum_{i=1}^N\left(T^{in}_i-1\right)^2/N}$, where we have used
$\left<T^{in}\right>=1$. For isothermal networks, which satisfy
$std(T^{in})=0$, the evolutionary dynamics under BD-B are equivalent
to the Moran process \cite{Lieberman05}.  Similarly, we define the
temperature for the outdegree by $T^{out}_i\equiv\sum_{j; (v_i,v_j)\in E}
\left(w_{ij}/\sum_l w_{lj}\right)$, which satisfies
$\left<T^{out}\right>=1$, and measure $std(T^{out})$.

\begin{figure}
\begin{center}
\includegraphics[height=6cm,width=6cm]{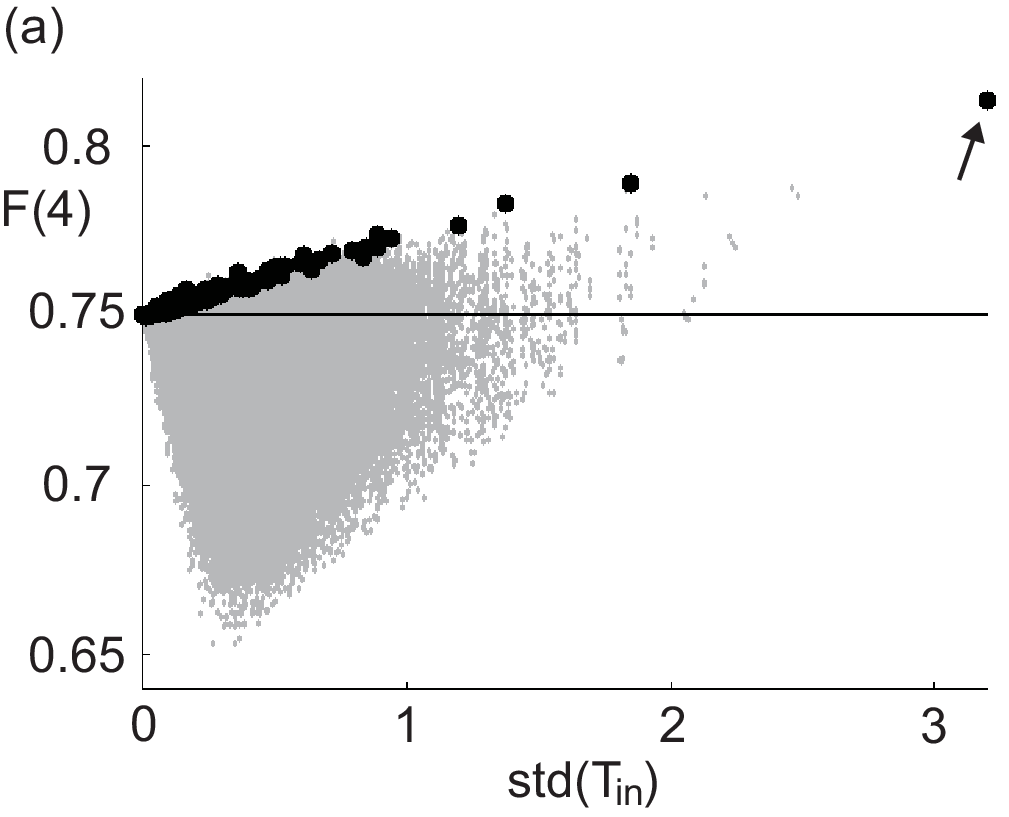}
\includegraphics[height=6cm,width=6cm]{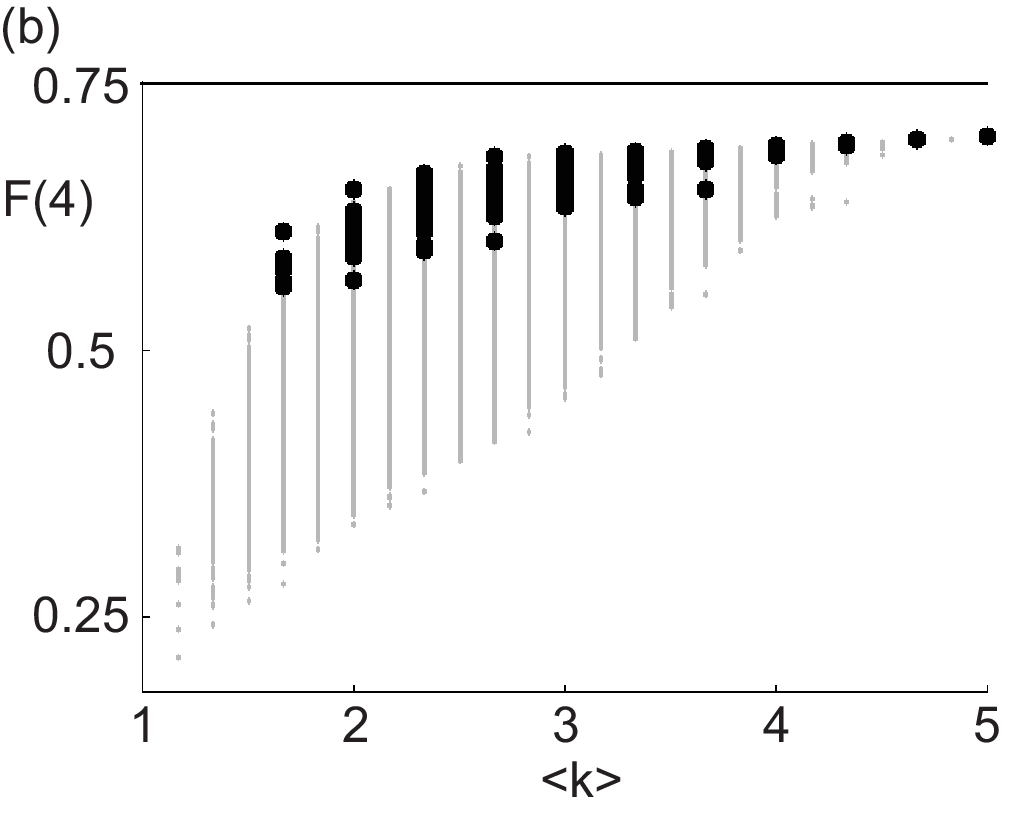}
\includegraphics[height=6cm,width=6cm]{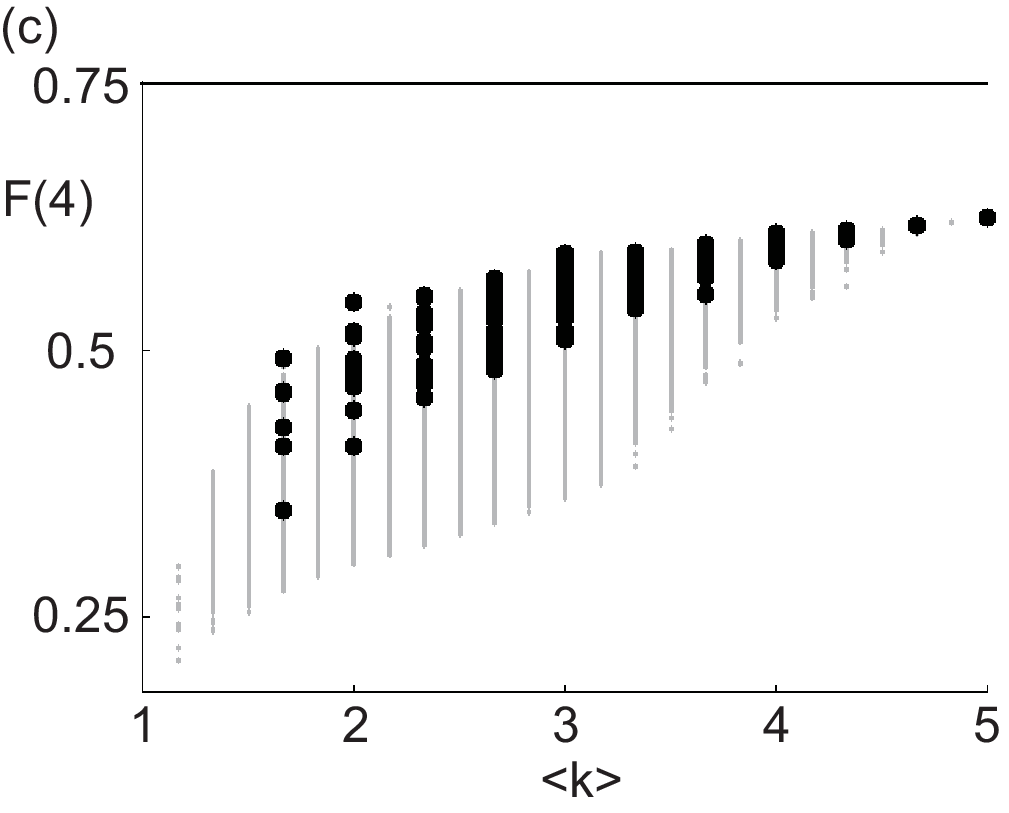}
\includegraphics[height=6cm,width=6cm]{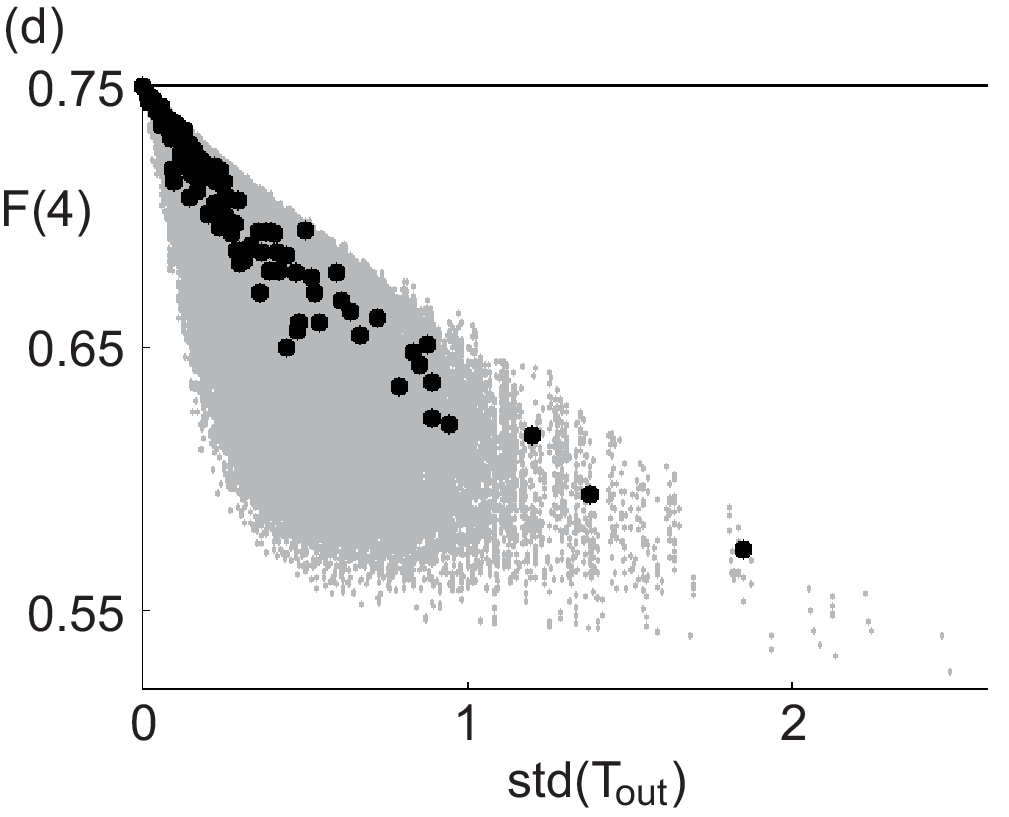}
\includegraphics[height=6cm,width=6cm]{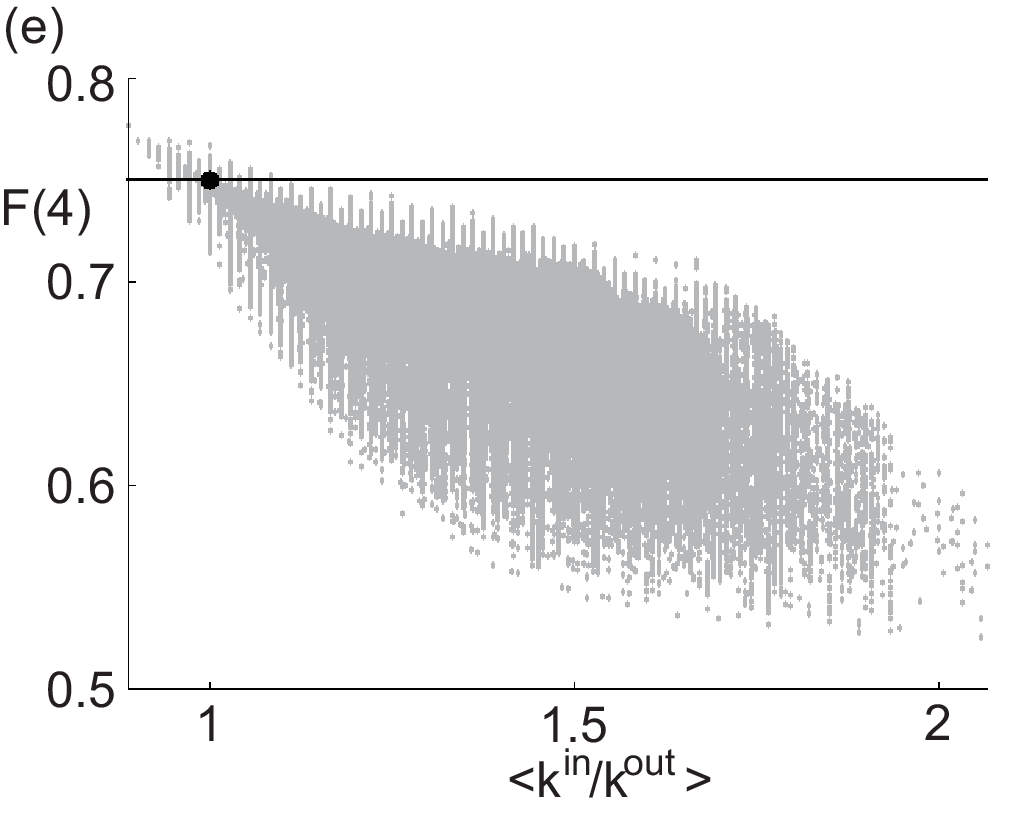}
\caption{Fixation probability $F(4)$ for the strongly
connected networks (gray dots) and for the undirected networks
(black circles) of size $N=6$. The line $F(4)=0.7502$ corresponds to
the Moran process.
(a) BD-B versus  $std(T^{in})$,
(b) BD-D versus $\left<k\right>$,
(c) DB-B versus $\left<k\right>$,
(d) DB-D versus $std(T^{out})$,
and (e) LD versus $\left<k^{in}/k^{out}\right>$.}
\label{fig:all N=6}
\end{center}
\end{figure}

Under BD-B, we have not found an order parameter that is strongly
correlated with $F(4)$, as listed in \TAB\ref{tab:regressor}.
For a visualization purpose, we plot
$F(4)$ against $std(T_{in})$ for all the networks
in \FIG\ref{fig:all N=6}(a).
Each gray dot in the figure corresponds to a network of size $N=6$.
Black circles are for undirected networks.
The Moran process with $N=6$ yields $F(4)=0.7502$, which
is shown by the solid line. Because $r=4>1$, networks with
$F(4)$ above (below) the solid line are amplifiers (suppressors).
Figure~\ref{fig:all
N=6}(a) suggests the following.
First, the undirected star (indicated by the arrow)
is by far the most amplifying
among all the networks. Most networks that yield large $F(4)$ are
variants of the star and have large $std(T_{in})$. Second, all the
undirected networks are amplifiers under BD-B, consistent with the
result that $F(r)$ for undirected uncorrelated networks
is given by \EQ\eqref{eq:F K} with
$K=\left<k\right>\left<k^{-1}\right> >1$
 \cite{Antal06prl,Sood08pre}. Indeed, $F(4)$ for undirected networks
with $N=6$ (black circles) are strongly correlated with
$\left<k\right>\left<k^{-1}\right>$ (data not shown). Third, a majority
of directed networks is suppressor.
The mean and the standard deviation of $F(4)$
based on all the networks and those based
on the undirected networks
compared in \TAB\ref{tab:rules compare} indicate a significant
difference.

Under BD-D, $\left<k\right>$ is
most strongly correlated with $F(4)$.
The values of $F(4)$ for
different networks are plotted against $\left<k\right>$ in
\FIG\ref{fig:all N=6}(b).  The complete graph, which
corresponds to $\left<k\right>=N-1=5$,
is the least suppressing, although it is nevertheless a
suppressor, as shown in \SEC\ref{sub:complete}.  All the networks are
suppressors.  Table~\ref{tab:rules compare} indicates
that the suppression is generally stronger under BD-D than under BD-B
and that directed networks are stronger suppressors than undirected
networks on an average.

Under DB-B also,
$\left<k\right>$ is most strongly correlated with $F(4)$.
Figure~\ref{fig:all N=6}(c), which shows
the relation between $F(4)$ and $\left<k\right>$, indicates that
the complete graph achieves the largest $F(4)$, that all the networks are
suppressors, and that directed networks are more
suppressing than undirected networks on an average. These tendencies
are similar to those for BD-D.

Under DB-D, $std(T^{out})$ is most strongly correlated with $F(4)$.
Equation~\eqref{eq:F
K} with $K=\left<k \right>^2/ \left<k^2\right>$, which was
derived for undirected
uncorrelated networks \cite{Antal06prl,Sood08pre}, approximates our
numerical results for the undirected networks (data not shown). However,
$std(T^{out})$ seems to be a better predictor of $F(4)$ than $\left<k
\right>^2/\left<\left(k^{in}\right)^2\right>$, $\left<k
\right>^2/\left<\left(k^{out}\right)^2\right>$, and $\left<k
\right>^2/\left<k^{in}k^{out}\right>$, probably because of the small
network size.  Figure~\ref{fig:all N=6}(d) indicates that 
$F(4)$ for the networks
with $std(T^{out})=0$ is equal to $F(4)$
for the Moran process, that all the other networks are
suppressors, and that directed networks are more suppressing than
undirected networks on average.

Under LD, $\left<k^{in}/k^{out}\right>$ is most strongly correlated
with $F(4)$. Evolutionary dynamics for undirected networks are
equivalent to the Moran process \cite{Antal06prl,Sood08pre}.
Figure~\ref{fig:all N=6}(e) indicates that
directed networks are generally more suppressing than
undirected networks, that
there are both amplifiers and suppressors
among directed networks with the majority being suppressors, and that
smaller $\left<k^{in}/k^{out}\right>$ tends to yield
larger $F(4)$.

\begin{figure}
\begin{center}
\includegraphics[height=3cm,width=4.5cm]{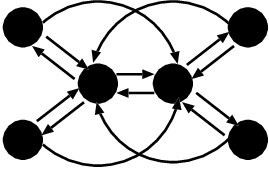}
\caption{The network with $N=6$ that yields the largest $F(4)$ under
LD.}
\label{fig:LD bestnet}
\end{center}
\end{figure}

The network that achieves the largest $F(4)$ is shown in
\FIG\ref{fig:LD bestnet}. Most networks that produce large $F(4)$ are
variants of this network. Because $\sum_{i=1}^N k_i^{in}
=\sum_{i=1}^N k_i^{out}$, a small value of
$\left<k^{in}/k^{out}\right>$ requires that $k^{in}<k^{out}$ holds for
many small-degree nodes and that $k^{in}>k^{out}$ holds for a
relatively small number of hubs. The network shown in
\FIG\ref{fig:LD bestnet} complies with this property.  In
\FIG\ref{fig:LD bestnet}, each of the two hubs links to the half of
the peripheral nodes, whereas each peripheral node links to both hubs.
If we merge the two hubs into one as an approximation, the network is
regarded as a weighted star with $a=2$ (\FIG\ref{fig:graphs}(d)).
The weighted star with $a=2$ is indeed an amplifier, as shown in
\SEC\ref{sub:star}.

Based on the numerical results for the five update rules, we claim the
following.  First, directed networks tend to be suppressors compared
to undirected networks regardless of the update rule.  Second, some
update rules (i.e., BD-D, DB-B, and DB-D) are much more suppressing
than the others (i.e., BD-B and LD). Particularly, the magnitude of the
amplification is ordered as: BD-B $>$ LD $>$ DB-D $>$ BD-D $>$ DB-B
(\TAB\ref{tab:rules compare}), which is consistent with the one for
the undirected star with small $N$ (\FIG\ref{fig:F simple}(f)). In a
single time step, the selection pressure operates on $N$ nodes for
BD-B, DB-D, and LD.  However, it operates on at most $N-1$ nodes for
BD-D and DB-B, because the node first selected for reproduction under
BD-D or for death under DB-B does not participate in the competition.
This is a main reason why BD-D and DB-B are strongly suppressing.  We
will corroborate these points by numerical simulations of large networks in
\SEC\ref{sub:large}.

\subsection{Large networks}\label{sub:large}

Because of the computational cost,
the exact numerical analysis performed in \SEC\ref{sub:N=6} is
feasible only for small networks.
In this section, we examine evolutionary
dynamics on larger networks by Monte Carlo simulations.

We use the Erd\"{o}s--R\'{e}nyi (ER) random graph, SF
networks, and their directed versions of different
sizes
(e.g. Albert and Barab\'{a}si, 2002; Newman, 2003).
We generate the undirected ER graph of mean degree
$\left<k\right>$ by connecting each pair of nodes with probability
$\left<k\right>/(N-1)$. The directed ER graph is generated by
connecting each ordered pair of nodes with probability
$2\left<k\right>/(N-1)$ so that
$\left<k^{in}\right>=\left<k^{out}\right>=\left<k\right>$. In both 
cases, the degree
distribution $p(k)$ follows a poisson distribution with mean
$\left<k\right>$: $p(k)\approx e^{-k/\left<k\right>}/\left<k\right>$.
For SF networks, we assume
$p(k)\propto k^{-3}$, ($\left<k\right>/2\le k$), where the
power-law exponent $3$ is an arbitrary choice. The SF network
represents the situation in which the degree is strongly
heterogeneous
\cite{Albert02,Newman03siam}.  To generate a SF network, we first
determine the degree of each node stochastically according to the
power-law 
distribution $p(k)$ with the restriction that the sum of the degree is
even. Then, we randomly add edges one by one so that the predetermined
degrees are respected at each node
\cite{Albert02,Newman03siam}.  For
undirected and directed SF networks, the added edges are undirected
and directed, respectively.  In each of the four network models, we
discard networks that are not strongly connected.

To calculate $F(4)$, we perform 2000 runs
for each initial
location of the type $A$ individual and count the
fraction of the $2000N$ runs in which 
the unanimity of type $A$ is reached.
This is a Monte Carlo 
realization of the fixation probability.  For a fixed $N$,
$\left<k\right>$, network model, and update rule,
we calculate the average and the standard
deviation of $F(4)$ based on 20 samples of networks.

\begin{figure}
\begin{center}
\includegraphics[height=4cm,width=4cm]{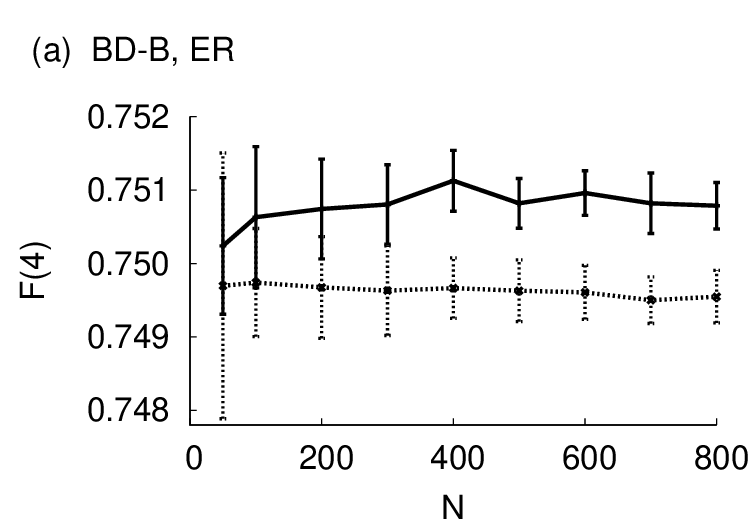}
\includegraphics[height=4cm,width=4cm]{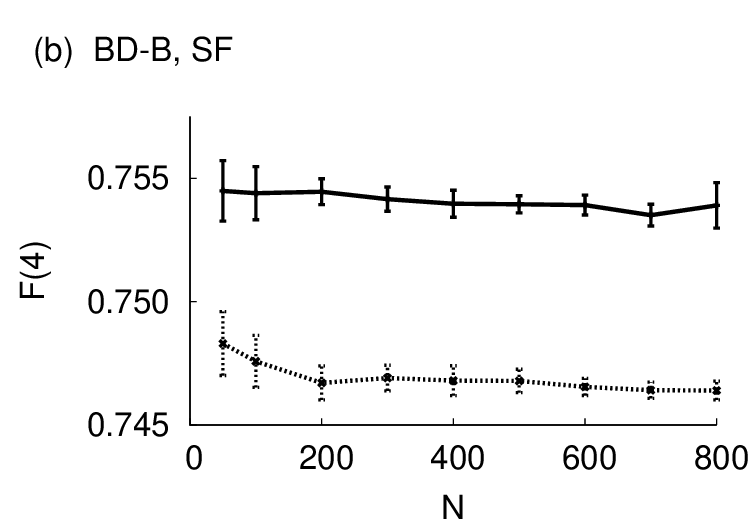}
\includegraphics[height=4cm,width=4cm]{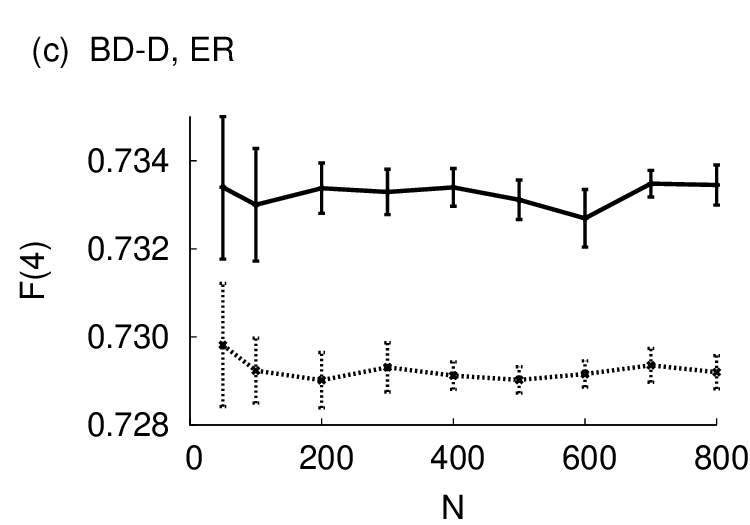}
\includegraphics[height=4cm,width=4cm]{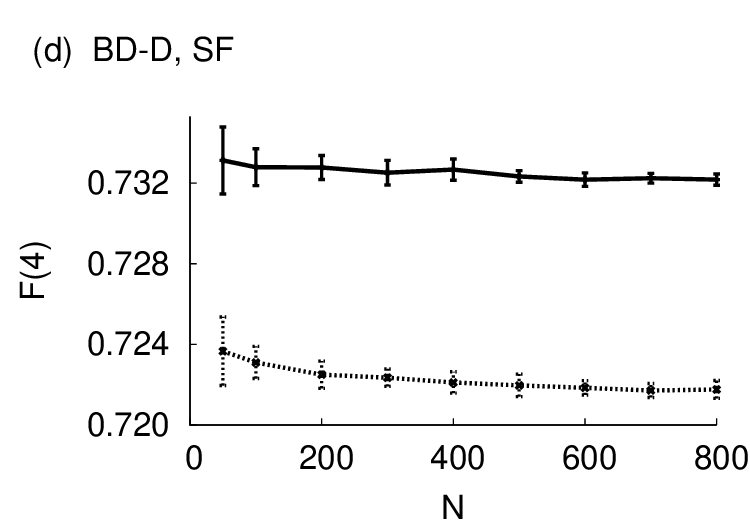}
\includegraphics[height=4cm,width=4cm]{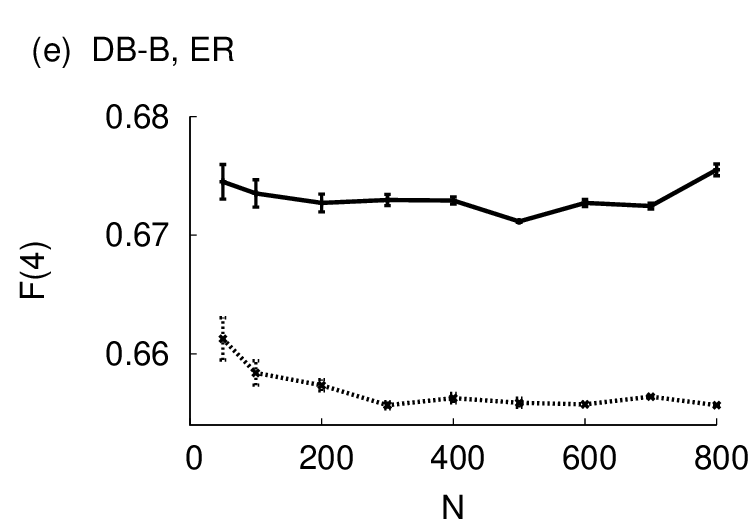}
\includegraphics[height=4cm,width=4cm]{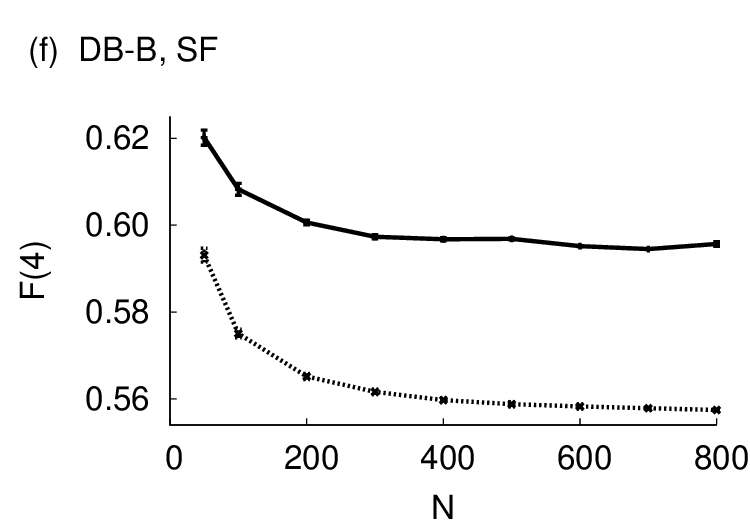}
\includegraphics[height=4cm,width=4cm]{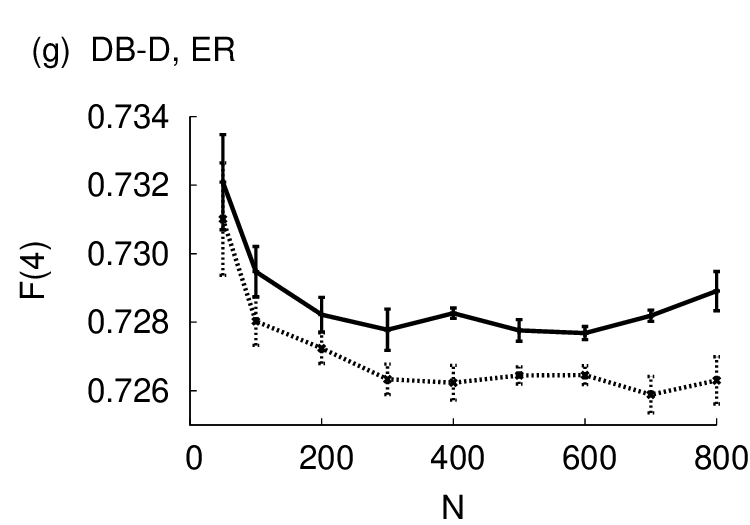}
\includegraphics[height=4cm,width=4cm]{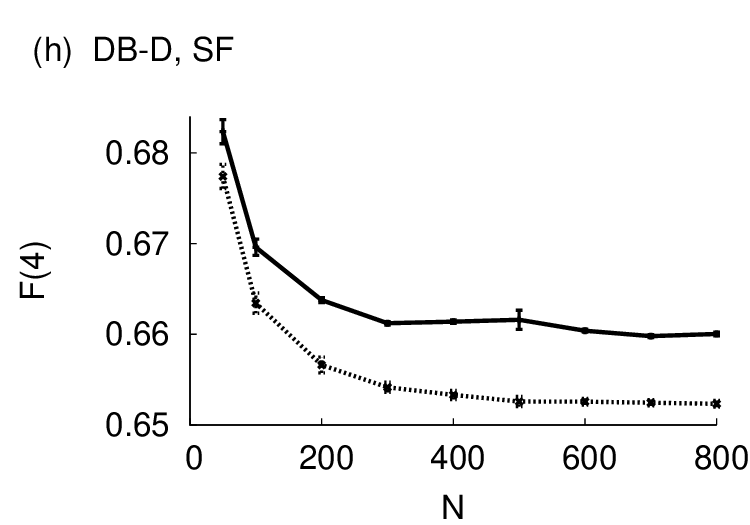}
\includegraphics[height=4cm,width=4cm]{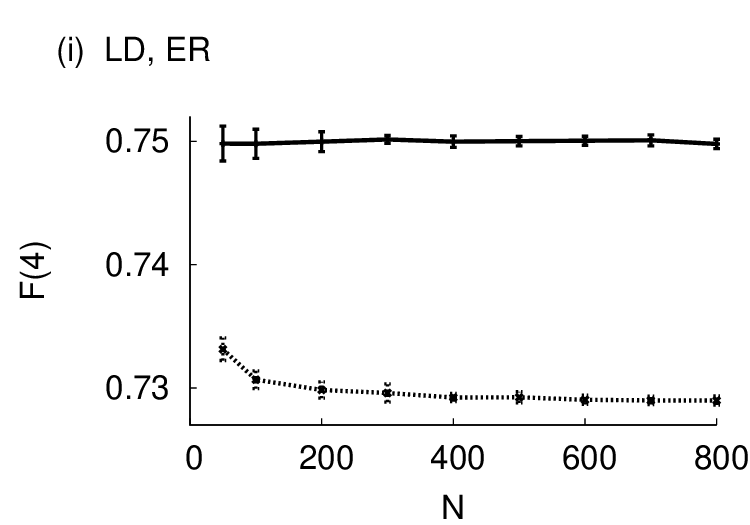}
\includegraphics[height=4cm,width=4cm]{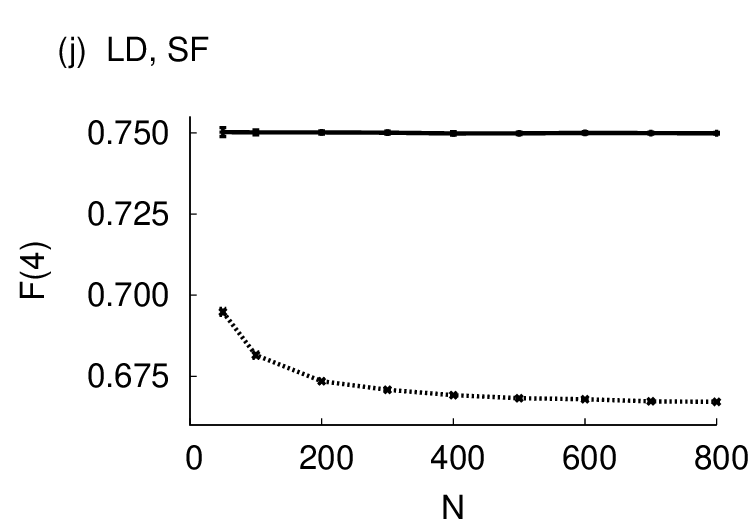}
\caption{Fixation probability $F(4)$ for large undirected networks
(solid lines) and directed networks (dashed lines) of different
size $N$. The update rules are (a, b) BD-B, (c, d) BD-D, (e, f) DB-B,
(g, h) DB-D, and (i, j) LD.  We use the ER random graphs in (a, c, e,
g, i) and the SF networks with $p(k)\propto k^{-3}$ in (b, d, f, h,
j). We set $\left<k\right>=10$. The Moran process yields $F(4)\approx
0.75$.}
\label{fig:F(4) large nets}
\end{center}
\end{figure}

For $\left<k\right>=10$,
$F(4)$ under BD-B is
plotted against $N$ for the ER and SF networks in
\FIGS\ref{fig:F(4) large nets}(a) and (b), respectively.
The standard deviation is shown by the error bars.
The SF networks are more amplifying than the ER
networks, consistent with the previous results
\cite{Lieberman05,Antal06prl,Sood08pre}. In
addition, undirected networks (solid lines)
are more amplifying than directed
networks (dashed lines), 
even for large $N$. The same tendencies are found for BD-D
(\FIGS\ref{fig:F(4) large nets}(c) and (d)), DB-B (\FIGS\ref{fig:F(4)
large nets}(e) and (f)), DB-D (\FIGS\ref{fig:F(4) large nets}(g) and
(h)), and LD (\FIGS\ref{fig:F(4) large nets}(i) and (j)).  This is a
strong indication that directionality of edges generally suppresses
the selection pressure.

\begin{figure}
\begin{center}
\includegraphics[height=4cm,width=4cm]{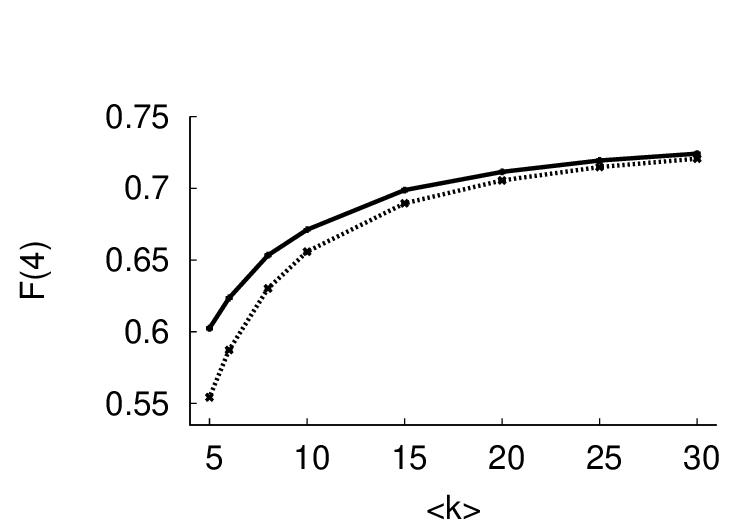}
\caption{Fixation probability 
$F(4)$ under DB-B for the undirected (solid line) and directed (dashed
line) ER network with different mean degrees.
We set $N=500$. The Moran process yields $F(4)\approx 0.75$.}
\label{fig:DB-B vary <k>}
\end{center}
\end{figure}

For each of the four types of networks used in \FIG\ref{fig:F(4)
large nets}, the strength of the selection pressure
is ordered as: BD-B $>$ BD-D,
LD $>$ DB-D $>$ DB-B. This order is consistent with the one for the
unweighted star with large $N$ (\FIG\ref{fig:F simple}(f)).
Compared to the results for the small networks shown in 
\SEC\ref{sub:N=6}, BD-D is much less suppressing in large networks, as
shown in \FIGS\ref{fig:F(4) large nets}(c) and (d).
This is a bit surprising because we have kept $\left<k\right>=10$ in
\FIG\ref{fig:F(4) large nets} so that the competition for
survival happens among roughly $\left<k\right>\ll N$ nodes per unit
time.
In contrast, DB-B remains strongly suppressing even for large $N$
(\FIGS\ref{fig:F(4) large nets}(e) and (f)).
An increase in $\left<k\right>$ with $N$ fixed
makes DB-B less
suppressing, as shown in \FIG\ref{fig:DB-B vary <k>}. This is consistent with the behavior of the ensemble of the small networks
(\FIG\ref{fig:all N=6}(c)).

\section{A theoretical explanation of the effect of directionality}\label{sec:two-node LD}

The finding that the directionality of the network suppresses the
selection pressure can be explained for LD on a simple network.
Consider a bidirectionally but asymmetrically connected 
two-node network. Edge ($v_1$, $v_2$) (from $v_1$ to $v_2$) has weight
unity, and edge ($v_2$, $v_1$) has weight $a$. Even though we
have formulated the evolutionary dynamics on unweighted networks, the
extension to the case of
weighted networks is straightforward (also refer to the
analysis of the weighted star in Appendix~B).
By enumerating the possible reproduction events per unit time, we
obtain
\begin{eqnarray}
F(AB) &=
\frac{rF(AA)+aF(BB)}{r+a} &= \frac{r}{r+a},
\label{eq:F(AB) LD}\\
F(BA) &=
\frac{raF(AA)+F(BB)}{ra+1} &= \frac{ra}{ra+1},
\label{eq:F(BA) LD}
\end{eqnarray}
where the first and the second arguments of $F$ are the types at $v_1$
and $v_2$, respectively, and 
$F$ is the fixation probability starting from that 
network state.
Note that $F(AA)=1$ and $F(BB)=0$.
Using \EQS~\eqref{eq:F(AB) LD} and \eqref{eq:F(BA) LD}, we obtain
\begin{equation}
F(r) = \frac{F(AB)+F(BA)}{2}=
\frac{1}{2}\left(\frac{r}{r+a}+\frac{ra}{ra+1}\right).
\end{equation}

For $r>1$, $F(r)$ as a function of $a$ takes the maximum at $a=1$. For
$r<1$, $F(r)$ takes the minimum at $a=1$. Therefore, the unweighted
network (i.e., $a=1$) is the most amplifying when we vary $a$. LD
on this unweighted network (i.e., $a=1$) is equivalent to the Moran process
with $N=2$. Accordingly, 
weighted networks (i.e., $a\neq 1$) are suppressors, and
the suppression is stronger as $a$ deviates from unity.

The applicability of the arguments above is beyond the two-node
network. Suppose that a network is divided into two modules with
homogeneous intramodular connectivity and that intermodular
connectivity is sparse and homogeneous, possibly with more edges from
one module to the other than the converse. Then the fixation
probability of a mutant on a particular node will depend only on the
module in which the initial mutant invades. In this case, we can
approximate the network by the two-node weighted network, where the
edge weights between the two aggregated nodes represent the gross
connectivity between the two modules.

\section{Discussion}

Motivated by the observation
that many real ecological and social networks 
underlying reproduction in
evolutionary dynamics are directed or weighted,
we have investigated the fixation probability on networks with
directed edges under five update rules. For undirected networks,
selection pressure relies on the network and the update
rule, which reproduces the previous results
\cite{Lieberman05,Antal06prl,Sood08pre}.  Our main conclusions in the
present paper are twofold.  

First, directionality of edges suppresses
the selection pressure for all the network types and the update rules
that we have dealt with. We have presented numerical evidence for
different sizes of networks (\SEC\ref{sec:numerical results}) and
simple analytical arguments (\SEC\ref{sec:two-node LD}) to support
this claim.
Note that many ecological and social situations in which evolutionary
dynamics take place can be modeled as directed or weighted
networks rather than undirected and unweighted
networks. Spreads of a type from one
habitat to another may be easier than in the other direction because of
heterogeneity in habitats and other geographical factors
(see \SEC\ref{sec:introduction} for references).
Accordingly,
fixation in real contact networks
may be less controlled by the values of fitness than in the corresponding
undirected networks
or well-mixed populations.

Second, the strength of the selection pressure depends on the update
rule for various networks, no matter whether edges are directed or not
(\SECS\ref{sec:simple} and \ref{sec:numerical results}).  This
conclusion extends the previous findings \cite{Antal06prl,Sood08pre}.
For BD-D, DB-B, and DB-D, no evolutionary amplifier has been found by
our exhaustive numerical analysis for small $N$ (\SEC\ref{sub:N=6}).
Moreover, DB-B is strongly suppressing compared to the other update
rules. In evolutionary game theory, DB-B
\cite{Nakamaru98jtb,Ohtsuki06nat,Ohtsuki06jtb} and its variants
\cite{Nowak92,Santos05prl} have commonly been used.  In games, the
fitness of a type is not constant, as assumed in the present paper,
but depends on neighbors' types.  Therefore, the results obtained in
this work are not immediately applicable to evolutionary
games. However, these previous results on evolutionary games might
considerably change under update rules less suppressing than DB-B, as
has been argued \cite{Nakamaru98jtb,Ohtsuki06nat,Ohtsuki06jtb}.

We have treated directed but unweighted networks for the sake of
clarity of the analysis. 
However, 
we believe that our conclusions for directed networks extend to
the case of weighted networks. This is partly supported by
the analysis of the two-node weighted network presented
in \SEC\ref{sec:two-node
LD}. In addition, if we quantize the edge weight and allow multiple
directed edges between two nodes, a weighted
network can be regarded as a directed unweighted network.

A unique point in our analysis is the use of very small networks with
$N=6$.  Small networks per se appear in many ecological contexts. For
example, the number of relevant habitats may not be very
large \cite{Gustafson96,Tischendorf00,Schick07}.  In addition, small
directed networks are likely to be structural and functional 
building blocks of large networks
\cite{Milo02,Itzkovitz03}.
According to the present analysis, 
the effect of the directionality of edges and that of
the update rule are mostly consistent between small
and large networks.
We believe that
the conclusions of the present paper apply to
real evolutionary dynamics in populations of various scales.

\section*{Acknowledgments}

N.M thanks Hisashi Ohtsuki for valuable discussions.
N.M acknowledges the support from the Grants-in-Aid for Scientific Research
 (Nos. 20760258, 20540382) from MEXT, Japan.

\renewcommand{\thesection}{\roman{section}}
\setcounter{section}{0}

\section*{Appendix A: Numerical procedure for small networks}\label{sec:procedure}

Here, we explain the methods for exactly calculating
the fixation probability for
strongly connected networks.  Because each node takes either type $A$
or $B$, there are $2^N$ possible states of the evolutionary dynamics.
We define the fixation probability for a state, which may have
multiple type-$A$ nodes, by the probability that type $A$ fixates
starting from that state.  The fixation probability for the all-$A$
state is 1, and that for the all-$B$ state is 0.

Given a network, an update rule, and $r$, the evolutionary
dynamics are equivalent to a nearest-neighbor random walk on the
$N$-dimensional hypercube comprising $2^N$ points.
In each time step, the
type at a node replaces the type at one of its neighbors in the original
evolutionary dynamics. Therefore, at most one node flips its type per
unit time. In terms of the random walk on the hypercube,
the walker moves to a neighboring state or does not move.
The random walk continues until either the all-$A$ state or
the all-$B$ state is reached.

\begin{figure}
\begin{center}
\includegraphics[height=4cm,width=4cm]{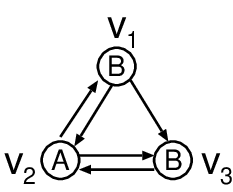}
\caption{A directed network with $N=3$.}
\label{fig:example-n3}
\end{center}
\end{figure}

Consider an example network shown in \FIG\ref{fig:example-n3}.
Denote the fixation probability for the state shown in 
\FIG\ref{fig:example-n3} by
$F(BAB)$, where the first, second, and the third arguments ($A$ or $B$)
correspond to the types at
$v_1$, $v_2$, and $v_3$, respectively. 
In a unit time, state $BAB$ may change to
$AAB$, $BBB$, or $BAA$. Otherwise, it does not change.
Under BD-B, for example, the fixation
probability satisfies
\begin{equation}
F(BAB) = \frac{1}{r+2}\frac{F(BBB)+F(BAB)}{2}+\frac{r}{r+2}\frac{F(AAB)+F(BAA)}{2}+\frac{1}{r+2}F(BBB).
\label{eq:example-bd-b}
\end{equation}
The first, second, and third terms in the RHS of
\EQ\eqref{eq:example-bd-b} represent the propagation of the type at
$v_1$, $v_2$, and $v_3$ to its neighbor, respectively. Noting the boundary
conditions $F(BBB)=0$ and $F(AAA)=1$, we can write
down the other five linear equations corresponding to the single-step
transition of $F(ABB)$, $F(BBA)$,
$F(AAB)$, $F(ABA)$, and
$F(BAA)$. The fixation probabilities
are obtained by solving the system of
the six linear equations. The fixation probability starting
from a single type-$A$ node is given by
$F(r)=\left[F\left(ABB\right)+F\left(BAB\right)
+F\left(BBA\right)\right]/3$.
Generally speaking, the fixation
probability for an $N$-node network is obtained by solving a system of
$2^N-2$ linear equations.
A standard method such as the Gauss elimination requires
$O\left(\left(2^N\right)^3\right)$ time of computation.

For small $N$, we can
enumerate all the possible networks \cite{Milo02,Itzkovitz03}.
There are 1047008 strongly connected directed
networks out of 1530843 weakly connected directed
networks with $N=6$. 
We cannot solve the fixation probability for $N>6$ 
due to the computational cost of
enumerating the directed
networks, but not due to the cost of
solving systems of $2^N-2$ linear equations.

\section*{Appendix B: Fixation probability for simple graphs}\label{sec:calculations}

In this appendix, we show detailed calculations of the fixation
probability for simple graphs. Following a standard procedure,
we map the evolutionary dynamics onto the discrete-time 
nearest-neighbor random walk on interval
$\{0,1,2,\ldots,N\}$. The position on the interval corresponds to
the number of type-$A$ individuals in a network of size $N$.
Positions 0 and $N$ are the 
only absorbing states. The type-$A$ individuals increase or
decrease at most by one per unit time.
The probability $F_m$ that the walker at position
$m$ eventually reaches position $N$ satisfies the following relations:
\begin{eqnarray}
F_0 &=& 0,\\
F_m &=& \alpha_m F_{m+1} + (1-\alpha_m-\beta_m)F_m
+ \beta_m F_{m-1},\quad (1\le m\le N-1)
\label{eq:Fm master}\\ 
F_N &=& 1,
\end{eqnarray}
for some $\alpha_m$ and $\beta_m$ ($1\le m\le N-1$).
Then, the fixation probability for a single mutant
is represented by \cite{Moran58,Ewensbook,Nowak06book}:
\begin{equation}
F_1 = \frac{1}{1+\sum^{N-1}_{m=1}\prod^m_{m^{\prime}=1}
\frac{\beta_{m^{\prime}}}{\alpha_{m^{\prime}}}}.
\label{eq:theorem}
\end{equation}

\subsection*{Complete graph}

Consider the complete graph with $N$ nodes.
The evolutionary dynamics under BD-B are equivalent
to the Moran process, and
\EQ\eqref{eq:Fm master} becomes
\begin{equation}
F_m=\frac{mr\frac{(N-m)F_{m+1}+(m-1)F_m}{N-1}
+
(N-m)\frac{(N-m-1)F_m+mF_{m-1}}{N-1}}
{mr+N-m}.
\end{equation}
Therefore, $\alpha_m=mr(N-m)/[(mr+N-m)(N-1)]$, 
$\beta_m=m(N-m)/[(mr+N-m)(N-1)]$, and
$\beta_m/\alpha_m=1/r$, which plugged into
\EQ\eqref{eq:theorem} leads to \EQ\eqref{eq:F moran}.
DB-D and LD also
yield $\beta_m/\alpha_m=1/r$ and hence the same result as that for
BD-B.

For BD-D, we obtain
\begin{equation}
F_m=\frac{m}{N}\frac{(N-m)F_{m+1}+\frac{m-1}{r}F_m}
{\frac{m-1}{r}+N-m}
+
\frac{N-m}{N}\frac{(N-1-m)F_m+\frac{m}{r}F_{m-1}}{\frac{m}{r}+N-1-m}.
\end{equation}
Therefore, 
\begin{equation}
\frac{\beta_m}{\alpha_m}=\frac{m-1+r(N-m)}
{r\left[m+r\left(N-1-m\right)\right]}.
\label{eq:gamma BD-D complete}
\end{equation}
Plugging \EQ\eqref{eq:gamma BD-D complete} into
\EQ\eqref{eq:theorem} leads to \EQ\eqref{eq:F BD-D complete}.

For DB-B, we obtain
\begin{equation}
F_m=\frac{m}{N}\frac{r(m-1)F_m+(N-m)F_{m-1}}{r(m-1)+N-m}
+
\frac{N-m}{N}\frac{rm F_{m+1}+(N-1-m)F_m}{rm+N-1-m}.
\end{equation}
Therefore,
\begin{equation}
\frac{\beta_m}{\alpha_m}=\frac{rm+N-1-m}
{r\left[r(m-1)+N-m\right]}.
\label{eq:gamma DB-B complete}
\end{equation}
Plugging \EQ\eqref{eq:gamma DB-B complete} into
\EQ\eqref{eq:theorem} leads to \EQ\eqref{eq:F DB-B complete}.

\subsection*{Undirected cycle}

Consider the undirected cycle with $N$ nodes. Under BD-D,
we obtain
\begin{eqnarray}
F_1 &=& \frac{1}{N}F_2+
\frac{2}{N}\frac{F_1+\frac{F_0}{r}}{1+\frac{1}{r}}
+\frac{N-3}{N}F_1,\\
F_m &=& \frac{2}{N}\frac{F_{m+1}+\frac{F_m}{r}}{1+\frac{1}{r}}
+\frac{2}{N}\frac{F_m+\frac{F_{m-1}}{r}}{1+\frac{1}{r}}
+\frac{N-4}{N}F_m,
\quad (2\le m\le N-2)\\
F_{N-1} &=& \frac{2}{N}\frac{F_N+\frac{F_{N-1}}{r}}{1+\frac{1}{r}}
+\frac{1}{N}F_{N-2}+\frac{N-3}{N}F_{N-1}.
\label{eq:BD-D cycle}
\end{eqnarray}
Therefore,
\begin{equation}
\frac{\beta_m}{\alpha_m}=\left\{
\begin{array}{ll}
\frac{2}{r+1}, & (m=1)\\
\frac{1}{r}, & (2\le m\le N-2)\\
\frac{r+1}{2r}. & (m=N-1)
\end{array}\right.
\label{eq:betaalpha BD-D cycle}
\end{equation}
Plugging \EQ\eqref{eq:betaalpha BD-D cycle} into
\EQ\eqref{eq:theorem} leads to \EQ\eqref{eq:F BD-D cycle}.

Under DB-B, we obtain
\begin{eqnarray}
F_1 &=& \frac{2}{N}\frac{rF_2+F_1}{r+1}+
\frac{1}{N}F_0 +\frac{N-3}{N}F_1,\\
F_m &=& \frac{2}{N}\frac{rF_{m+1}+F_m}{r+1}
+\frac{2}{N}\frac{rF_m+F_{m-1}}{r+1}
+\frac{N-4}{N}F_m,
\quad (2\le m\le N-2)\\
F_{N-1} &=& \frac{1}{N}F_N
+\frac{2}{N}\frac{rF_{N-1}+F_{N-2}}{r+1}+\frac{N-3}{N}F_{N-1}.
\label{eq:DB-B cycle}
\end{eqnarray}
Therefore,
\begin{equation}
\frac{\beta_m}{\alpha_m}=\left\{
\begin{array}{ll}
\frac{r+1}{2r}, & (m=1)\\
\frac{1}{r}, & (2\le m\le N-2)\\
\frac{2}{r+1}. & (m=N-1)
\end{array}\right.
\label{eq:betaalpha DB-B cycle}
\end{equation}
Plugging \EQ\eqref{eq:betaalpha DB-B cycle} into
\EQ\eqref{eq:theorem} leads to \EQ\eqref{eq:F DB-B cycle}.

\subsection*{Star}

\begin{figure}
\begin{center}
\includegraphics[height=4cm,width=4cm]{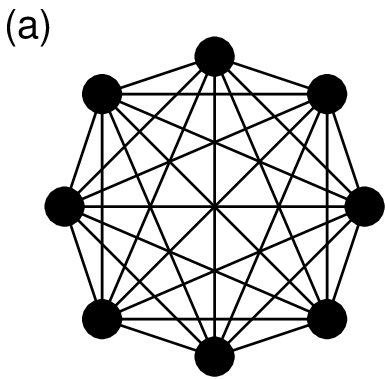}
\includegraphics[height=4cm,width=4cm]{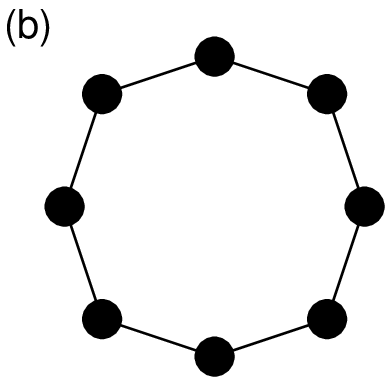}
\includegraphics[height=4cm,width=4cm]{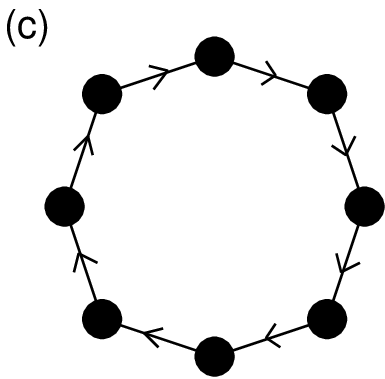}
\includegraphics[height=4cm,width=4cm]{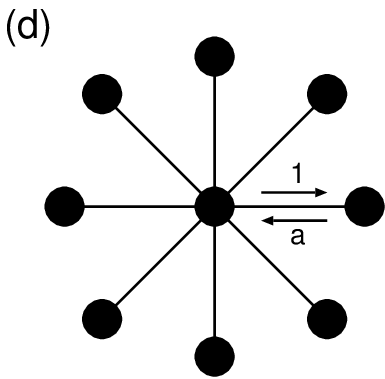}
\caption{(a) Complete graph, (b) undirected cycle, (c) directed
cycle, and (d) weighted star.}
\label{fig:graphs}
\end{center}
\end{figure}

Consider the undirected star with $N$ nodes. 
To calculate the fixation probability for BD-B and LD simultaneously,
it is convenient to introduce the edge weight.
We assume that an edge from the hub to a leaf is endowed with weight unity 
and one from a leaf to the hub is endowed with weight $a$ (\FIG\ref{fig:graphs}(d)).
The edge weight works as a multiplicative factor
to the probability that this edge is
chosen for reproduction, which is explained in \SEC\ref{sec:model}.
Denote by $F_{m,A}$ ($F_{m,B}$)
($0\le m\le N-1$) the fixation probability 
of type $A$ when there are $m$ type-$A$ leaves
and the hub is
occupied by type $A$ ($B$).

BD-B for the unweighted star is equivalent to 
LD with selection on the birth, which is actually the ordinary LD,
for the weighted star with $a=N-1$.
DB-D for the unweighted star is equivalent to
LD with selection on the death, which is again the ordinary LD, for the 
weighted star with $a=1/(N-1)$.
Therefore, we calculate the fixation probability for the weighted
star under LD. Interpreting LD as the LD with selection on the birth,
we obtain
\begin{eqnarray}
F_{m,A} &=& \frac{rma F_{m,A}+(N-1-m)a F_{m,B}+rm F_{m,A}+r(N-1-m)
F_{m+1,A}}{rma+(N-1-m)a+r(N-1)},
\label{eq:star b m1-1}\\
&&(0\le m\le N-2)\nonumber\\
F_{m,B} &=& \frac{rmaF_{m,A}+(N-1-m)aF_{m,B}+mF_{m-1,B}+(N-1-m)F_{m,B}}
{rma+(N-1-m)a+(N-1)}.
\label{eq:star b m0-1}\\
&&(1\le m\le N-1)\nonumber
\end{eqnarray}
\EQS\eqref{eq:star b m1-1} and \eqref{eq:star b m0-1}, respectively,
lead to
\begin{eqnarray}
(r+a)F_{m,A} &= aF_{m,B}+rF_{m+1,A},& (0\le m\le N-2)
\label{eq:star b m1-2}\\
(ra+1)F_{m,B} &= raF_{m,A}+F_{m-1,B}.& (1\le m\le N-1)
\label{eq:star b m0-2}
\end{eqnarray}
By combining \EQS\eqref{eq:star b m1-2} and 
\eqref{eq:star b m0-2}, we
obtain
\begin{equation}
(r^2 a+2r+a)F_{m,A} - r(ra+1)F_{m+1,A} - (r+a)F_{m-1,A}=0.
\quad (1\le m\le N-2)
\label{eq:star b 3}
\end{equation}
The solution to \EQ\eqref{eq:star b 3} that satisfies
$(r+a)F_{0,A}=rF_{1,A}$, which comes from
\EQ\eqref{eq:star b m1-2} with $m=0$ and $F_{0,B}=0$,
and $F_{N-1,A}=1$ is
\begin{equation}
F_{m,A}=\frac{\frac{r(r+a)}{ra+1}-\left(\frac{r+a}{r(ra+1)}\right)^m}
{\frac{r(r+a)}{ra+1}-\left(\frac{r+a}{r(ra+1)}\right)^{N-1}}.
\label{eq:star b m1-final}
\end{equation}
Plugging \EQ\eqref{eq:star b m1-final} into \EQ\eqref{eq:star b m1-2}
leads to
\begin{equation}
F_{m,B}=\frac{\frac{r(r+a)}{ra+1}\left[1
-\left(\frac{r+a}{r(ra+1)}\right)^m\right]}
{\frac{r(r+a)}{ra+1}-\left(\frac{r+a}{r(ra+1)}\right)^{N-1}}.
\label{eq:star b m0-final}
\end{equation}
The fixation probability for a single mutant is represented by
\begin{equation}
F(r)=\frac{(N-1)F_{1,B}+F_{0,A}}{N}=
\frac{1-\frac{N-1}{N}\frac{r+a}{r(ra+1)}-\frac{1}{N}\frac{ra+1}{r(r+a)}}
{1-r^{-N}\left(\frac{r+a}{ra+1}\right)^{N-2}}.
\end{equation}


For BD-D on the unweighted star, we obtain
\begin{eqnarray}
F_{m,A} &=& \frac{\frac{\frac{m}{r}}{\frac{m}{r}+N-1-m}F_{m,A}
+\frac{N-1-m}{\frac{m}{r}+N-1-m}F_{m+1,A}+
mF_{m,A}+(N-1-m)F_{m,B}
}{N},
\label{eq:star BD-D m1-1}\\
&&(0\le m\le N-2)\nonumber\\
F_{m,B} &=& \frac{\frac{\frac{m}{r}}{\frac{m}{r}+N-1-m}F_{m-1,B}
+\frac{N-1-m}{\frac{m}{r}+N-1-m}F_{m,B}
+mF_{m,A}+(N-1-m)F_{m,B}}{N}.
\label{eq:star BD-D m0-1}\\
&&(1\le m\le N-1)\nonumber
\end{eqnarray}
Equations~\eqref{eq:star BD-D m1-1} 
and \eqref{eq:star BD-D m0-1}, respectively,
lead to
\begin{eqnarray}
\left[r\left(N-m\right)+m\right]F_{m,A} 
&=& \left[m+r\left(N-1-m\right)\right]F_{m,B}+rF_{m+1,A},
\label{eq:star BD-D m1-2}\\
&& (0\le m\le N-2)\nonumber\\
\left[m+1+r\left(N-1-m\right)\right]F_{m,B} 
&=& \left[m+r\left(N-1-m\right)\right]F_{m,A}+F_{m-1,B}.
\label{eq:star BD-D m0-2}\\
&& (1\le m\le N-1)\nonumber
\end{eqnarray}
By combining \EQS\eqref{eq:star BD-D m1-2} and
\eqref{eq:star BD-D m0-2}, we obtain
\begin{equation}
F_{m+1,A}-F_{m,A} =\frac{\left[m+r\left(N-1-m\right)\right]
\left[m-1+r\left(N+1-m\right)\right]}
{r\left[m+1+r\left(N-1-m\right)\right]
\left[m-1+r\left(N-m\right)\right]}\left(F_{m,A}-F_{m-1,A}\right),
\quad (1\le m\le N-2)
\label{eq:star BD-D 3}
\end{equation}
which yields
\begin{eqnarray}
&&F_{N-1,A}-F_{0,A}\nonumber\\
&=& \sum^{N-2}_{m=0}\prod^m_{m^{\prime}=1}
\frac{\left[m^{\prime}+r\left(N-1-m^{\prime}\right)\right]
\left[m^{\prime}-1+r\left(N+1-m^{\prime}\right)\right]}
{r\left[m^{\prime}+1+r\left(N-1-m^{\prime}\right)\right]
\left[m^{\prime}-1+r\left(N-m^{\prime}\right)\right]}
\left(F_{1,A}-F_{0,A}\right)\nonumber\\
&=& \sum^{N-2}_{m=0}\frac{N\left[m+r\left(N-1-m\right)\right]
\left[1+r\left(N-1\right)\right]}
{r^m\left(N-1\right)\left[m+1+r\left(N-1-m\right)\right]
\left[m+r\left(N-m\right)\right]}
\left(F_{1,A}-F_{0,A}\right)\nonumber\\
&=& \frac{rN\left[1+r\left(N-1\right)\right]}
{(N-1)(r-1)} \sum^{N-2}_{m=0}
\left[\frac{r^{-m}}{m+r\left(N-m\right)}
-\frac{r^{-(m+1)}}{m+1+r\left(N-1-m\right)}
\right]\left(F_{1,A}-F_{0,A}\right)\nonumber\\
&=& \frac{\left[1+r\left(N-1\right)\right]}
{(N-1)(r-1)} 
\left[1- \frac{r^{-N+2}N}{N-1+r}\right]
\left(F_{1,A}-F_{0,A}\right).
\label{eq:star BD-D 4}
\end{eqnarray}
By substituting $F_{1,A}=NF_{0,A}$, which is derived by setting $m=0$ in
\EQ\eqref{eq:star BD-D m1-2}, and $F_{N-1,A}=1$
into \EQ\eqref{eq:star BD-D 4},
we obtain
\begin{equation}
F_{0,A}=\frac{1}{1+\frac{1+r(N-1)}{r-1}
\left(1-\frac{r^{-N+2}N}{(N-1+r)}\right)}.
\label{eq:star BD-D m01-final}
\end{equation}
By setting $m=1$ in \EQ\eqref{eq:star BD-D m0-2} and
using $F_{0,B}=0$, we derive
\begin{equation}
F_{1,B}=\frac{\left[1+r\left(N-2\right)\right]N}{2+r(N-2)}F_{0,A}.
\end{equation}
Therefore, the fixation probability 
$F(r)=[(N-1)F_{1,B}+F_{0,A}]/N$ is given by
\EQ\eqref{eq:star BD-D final}.

For DB-B on the unweighted star, we obtain
\begin{eqnarray}
F_{m,A} &=& \frac{mF_{m,A}+(N-1-m)F_{m+1,A}+
\frac{rm F_{m,A}+(N-1-m)F_{m,B}}{rm+N-1-m}}{N},
\label{eq:star DB-B m1-1}\\
&&(0\le m\le N-2)\nonumber\\
F_{m,B} &=& \frac{mF_{m-1,B}+(N-1-m)F_{m,B}
+\frac{rm F_{m,A}+(N-1-m)F_{m,B}}{rm+N-1-m}}{N}.
\label{eq:star DB-B m0-1}\\
&&(1\le m\le N-1)\nonumber
\end{eqnarray}
\EQS\eqref{eq:star DB-B m1-1} and \eqref{eq:star DB-B m0-1}, respectively,
lead to
\begin{eqnarray}
(rm+N-m)F_{m,A} &=& F_{m,B}+(rm+N-1-m)F_{m+1,A},
\label{eq:star DB-B m1-2}\\
&& (0\le m\le N-2)\nonumber\\
\left[r\left(m+1\right)+N-1-m\right]F_{m,B}
&=& rF_{m,A}+(rm+N-1-m)F_{m-1,B}.
\label{eq:star DB-B m0-2}\\
&& (1\le m\le N-1)\nonumber
\end{eqnarray}
By combining \EQS\eqref{eq:star DB-B m1-2} and
\eqref{eq:star DB-B m0-2}, we obtain
\begin{equation}
F_{m+1,A}-F_{m,A}=\frac{\left[r\left(m-1\right)+N+1-m\right]}
{\left[r\left(m+1\right)+N-1-m\right]}
\left(F_{m,A}-F_{m-1,A}\right),
\quad (1\le m\le N-2)
\label{eq:star DB-B 3}
\end{equation}
which yields
\begin{eqnarray}
F_{N-1,A}-F_{0,A} &=&
\sum^{N-2}_{m=0}\prod^m_{m^{\prime}=1}
\frac{\left[r\left(m^{\prime}-1\right)+N+1-m^{\prime}\right]}
{\left[r\left(m^{\prime}+1\right)+N-1-m^{\prime}\right]}
\left(F_{1,A}-F_{0,A}\right)
\nonumber\\
&=& \frac{(N+r-1)(N-1)}{rN-r+1}
\left(F_{1,A}-F_{0,A}\right).
\label{eq:star DB-B 4}
\end{eqnarray}
By substituting $F_{1,A}=NF_{0,A}/(N-1)$, which is derived by setting
$m=0$ in \EQ\eqref{eq:star DB-B m1-2}, and $F_{N-1,A}=1$ into
\EQ\eqref{eq:star DB-B 4}, we obtain
\begin{equation}
F_{0,A}=\frac{1}{1+\frac{N+r-1}{rN-r+1}}.
\label{eq:star DB-B m01-final}
\end{equation}
By setting $m=1$
in \EQ\eqref{eq:star DB-B m0-2} and using
$F_{0,B}=0$, we derive
\begin{equation}
F_{1,B}=\frac{rN}{(N-1)(N+2r-2)}F_{0,A}.
\end{equation}
Therefore, the fixation probability 
$F(r)=[(N-1)F_{1,B}+F_{0,A}]/N$ is given by
\EQ\eqref{eq:star DB-B final}.

\newpage
\clearpage

\clearpage

\clearpage

\clearpage

\clearpage

\clearpage

\clearpage

\clearpage
\newpage

\begin{table}
\begin{center}
\caption{Correlation coefficient between
$F(4)$ and the order parameters of the network.}
\begin{tabular}{|p{2.7cm}|p{1.4cm}|p{1.4cm}|p{1.4cm}|p{1.4cm}|p{1.4cm}|}
\hline
Parameters & BD-B & BD-D & DB-B & DB-D & LD \\ \hline
$\left<k\right>$ & 0.2196 & 0.7719 & 0.7905 &
0.3791 & 0.2179\\ \hline
$\left<k\right>\left<1/k^{in}\right>$ & -0.3183 &
-0.2639 & -0.4478 & -0.3816 & -0.1892\\ \hline
$\left<k\right>\left<1/k^{out}\right>$ & 
-0.3486 & -0.6495 & -0.5415 & -0.7237 & -0.6993\\ \hline
$\left<k\right>^2/\left<\left(k^{in}\right)^2\right>$ & 
0.2983 & 0.3526 & 0.5493 & 0.4311 & 0.2046\\ \hline
$\left<k\right>^2/\left<\left(k^{out}\right)^2\right>$ &
0.3392 & 0.7205 & 0.6395 & 0.7496 & 0.6921\\ \hline
$\left<k\right>^2/\left<k^{in}k^{out}\right>$ &
-0.4461 & -0.2715 & -0.3130 & -0.1665 & -0.6770\\ \hline
$\left<k^{out}/k^{in}\right>$ &
-0.4640 & -0.3111 & -0.4824 & -0.3716 & -0.5048\\ \hline
$\left<k^{in}/k^{out}\right>$ &
-0.5261 & -0.6143 & -0.5369 & -0.5978 & -0.8557\\ \hline
$std(T^{in})$ & -0.3628 & -0.4664 &
-0.6233 & -0.7581 & -0.3613\\ \hline
$std(T^{out})$ & -0.3790 & -0.5965 &
-0.6510 & -0.8505 & -0.4935\\ \hline
\end{tabular}
\label{tab:regressor}
\end{center}
\end{table}

\clearpage
\newpage

\begin{table}
\begin{center}
\caption{Statistics of $F(4)$
for the networks with $N=6$.}
\begin{tabular}{|p{1.3cm}|p{3cm}|p{3cm}|}\hline
update & all networks & undirected only\\
rule & (ave $\pm$ std) & (ave $\pm$ std)\\ \hline
BD-B & 0.7422 $\pm$ 0.0107 & 0.7575 $\pm$ 0.0090\\ \hline
BD-D & 0.5864 $\pm$ 0.0552 & 0.6489 $\pm$ 0.0336\\ \hline
DB-B & 0.4806 $\pm$ 0.0485 & 0.5367 $\pm$ 0.0485\\ \hline
DB-D & 0.6986 $\pm$ 0.0317 & 0.7033 $\pm$ 0.0411\\ \hline
LD & 0.7088 $\pm$ 0.0308 & 0.7502 $\pm$ 0\\ \hline
\end{tabular}
\label{tab:rules compare}
\end{center}
\end{table}

\end{document}